\documentstyle[preprint,aps,floats]{revtex}
\input epsf
\def\GeV{{\rm\,GeV}}
\def\TeV{{\rm\,TeV}}
\def\lt{\lambda_{\rm t}}
\def\lb{\lambda_{\rm b}}
\def\ltau{\lambda_{\rm \tau}}
\def\lG{\lambda_{\rm G}}
\def\beq{\begin{equation}}
\def\eeq{\end{equation}}
\def\bea{\begin{eqnarray}}
\def\eea{\end{eqnarray}}
\def\msbar{\overline{\rm MS}}
\def\drbar{\overline{\rm DR}}
\def\sq{\tilde{q}}
\def\sl{\tilde{\ell}}
\def\hi{\tilde{h}}
\def\wi{\tilde{w}}
\def\bi{\tilde{b}}
\def\gl{\tilde{g}}
\def\tr{\tilde{t}_R}
\def\tl{\tilde{t}_L}
\def\st{\tilde{t}}
\def\br{\tilde{b}_R}
\def\bl{\tilde{b}_L}
\def\sb{\tilde{b}}
\def\taur{\tilde{\tau}_R}
\def\taul{\tilde{\tau}_L}
\def\PQ{${\cal PQ}$}
\def\R{${\cal R}$}
\def\epsr{\epsilon_{\cal R}}
\def\epspq{\epsilon_{\cal PQ}}

\begin{document}
\draft
\preprint{\vbox{\hbox{LBL-33997}
            \hbox{UCB-PTH-93/15}
            \hbox{hep-ph/9306309}
            \hbox{June 1993}
	    \hbox{Rev.: March 1994}}}
\title{The Top Quark Mass in Supersymmetric SO(10) Unification}
\author{Lawrence J.~Hall$^{\dagger\ddagger}$\cite{lhaddr}, {
Riccardo Rattazzi}$^{\dagger}$\cite{rraddr} and {Uri
Sarid}$^{\dagger}$\cite{usaddr}}

\address{
$^{\dagger}$Theoretical Physics Group, 50A/3115, Lawrence Berkeley
Laboratory,\\
 1 Cyclotron Road, Berkeley, California 94720 \\
%\vskip .2in \\
$^{\ddagger}$Physics Department, University of California, Berkeley,
California 94720
}

\maketitle

\begin{abstract}
The successful prediction of the weak mixing angle suggests that the
effective theory beneath the grand unification scale is the minimal
supersymmetric standard model (MSSM) with just two Higgs doublets.
If we
further assume that the unified gauge group contains SO(10),
that the two light Higgs doublets lie mostly in a single
irreducible SO(10) representation, and that the $t$, $b$ and $\tau$
masses originate in renormalizable Yukawa interactions of the form
$\underline{16}_3\,{\cal O}\,\underline{16}_3$, then also the top
quark mass can be predicted in terms of the MSSM parameters.
To compute $m_t$ we present a precise
analytic approximation to the solution of the 2-loop renormalization
group equations, and study supersymmetric and
GUT threshold corrections and the input value of the $b$ quark
mass. The large ratio of top to bottom quark masses derives
from a large ratio, $\tan\beta$, of Higgs vacuum expectation values.
We point out that when $\tan\beta$ is large, so are certain
corrections to the $b$ quark mass prediction, unless a particular
hierarchy exists in the parameters of the model. With such a
hierarchy, which may result from approximate symmetries, the top
mass prediction depends only weakly on the spectrum. Our
results may be applied to any supersymmetric SO(10) model
as long as $\lambda_t\simeq\lambda_b\simeq\lambda_\tau$ at the GUT
scale and there are no intermediate mass scales in the desert.
\end{abstract}

\narrowtext

%THIS PAGE (PAGE ii) CONTAINS THE LBL DISCLAIMER
%TEXT SHOULD BEGIN ON NEXT PAGE (PAGE 1)
\renewcommand{\thepage}{\roman{page}}
\setcounter{page}{2}
\mbox{ }

\vskip 1in

\begin{center}
{\bf Disclaimer}
\end{center}

\vskip .2in

\begin{scriptsize}
\begin{quotation}
This document was prepared as an account of work sponsored by the
United
States Government.  Neither the United States Government nor any
agency
thereof, nor The Regents of the University of California, nor any of
their
employees, makes any warranty, express or implied, or assumes any
legal
liability or responsibility for the accuracy, completeness, or
usefulness
of any information, apparatus, product, or process disclosed, or
represents
that its use would not infringe privately owned rights.  Reference
herein
to any specific commercial products process, or service by its trade
name,
trademark, manufacturer, or otherwise, does not necessarily
constitute or
imply its endorsement, recommendation, or favoring by the United
States
Government or any agency thereof, or The Regents of the University
of
California.  The views and opinions of authors expressed herein do
not
necessarily state or reflect those of the United States Government
or any
agency thereof of The Regents of the University of California and
shall
not be used for advertising or product endorsement purposes.
\end{quotation}
\end{scriptsize}

\vskip 2in

\begin{center}
\begin{small}
{\it Lawrence Berkeley Laboratory is an equal opportunity employer.}
\end{small}
\end{center}

\newpage
\renewcommand{\thepage}{\arabic{page}}
\setcounter{page}{1}
%THIS IS PAGE 1 (INSERT TEXT OF REPORT HERE)

\section{Introduction}

The standard model of particle physics is extraordinarily
successful,
describing all known properties of the elementary particles in terms
of just 18
free parameters. Nevertheless, the model leaves so many questions
unanswered
that numerous ideas and speculations leading toward a more
fundamental
theory have been developed. While there is no hard evidence to
support any of
these speculations, experiments have provided some hints. In
particular, the
only parameter of the standard model that has been successfully
predicted with
a high level of significance is the weak mixing angle, which is a
prediction of
supersymmetric grand unified theories (SUSY GUTs) \cite{ssq}.

While this is just a single
parameter, the excellent agreement of data with the simplest SUSY
GUT
suggests that it is worthwhile to pursue other predictions of
similar
SUSY GUTs. This is harder than it sounds. The reason is that the
weak mixing
angle has a unique status within these
theories: it is the only parameter which can be predicted
by knowing only the gauge group structure of the model. In fact all
one needs
to know \cite{caveat} is that the gauge group is, or breaks to,
SU(5)
\cite{gg}. All other predictions require additional, model-dependent
information
about the theory. A good example of this is the proton decay rate:
it can only
be computed after making an assumption about the spectrum of the
superheavy
colored states. It is a very interesting quantity, since the
simplest possible
structure for this superheavy spectrum gives a decay rate that may
well be
accessible to planned experimental searches. Nevertheless, only
minor changes
in the theory can lead to very large suppression factors in the
rate.

A potentially copious source of predictions is the flavor sector,
responsible for the
quark and lepton masses and mixings. An early success of GUTs was
the prediction of the bottom quark to tau lepton mass ratio,
$m_b/m_\tau$ \cite{ceg}; however, several considerations make this
less impressive
than the weak mixing angle prediction. First and foremost is the low
numerical
significance of the $m_b/m_\tau$ prediction: it cannot be made with
much accuracy as long as the strong coupling $\alpha_3$ and the top
quark Yukawa coupling (which cannot be neglected for a heavy top)
are not known very well, and it cannot be compared accurately with
experiment without better knowledge of $m_b$. This low
significance is especially troubling given that the theory is
predicting only
one of the 13 independent flavor parameters of the standard model.
Nevertheless, with a heavy top quark, an acceptable value of
$m_b/m_\tau$ {\it requires} more than one light Higgs
doublet---which provides another interesting hint pointing towards
low-energy supersymmetry \cite{twohiggs}. Finally,
the simplest flavor sector which leads to this prediction
immediately fails
when extended to the lighter generations: $m_s/m_d = m_\mu/m_e$ is
unacceptable. To overcome these objections it is
necessary to construct a more complicated flavor sector of the grand
unified
theory. Here there is a delicate balance: more structure requires
further
assumptions, but these are perhaps justified if there are additional
predictions. This approach was developed long ago \cite{gj,gn,hrr}
and has
received considerable attention recently \cite{latetextures}.
Using the full power of the grand unified group SO(10) \cite{soten}
it is
possible to obtain predictions for 7 of the 13 flavor parameters.
Further
development of the flavor sector can also lead to predictions for
neutrino
masses \cite{hrr,neutmass}. Despite these successes, one still has
to admit that
these schemes are based on the hope that the flavor sector at the
grand unified
scale is particularly simple: the quark and lepton masses must
originate in
just a few grand unified interactions. If there are many such
interactions the
predictions are lost. This is a particularly acute problem for the
lighter
generations.
The smallness of these masses can be understood if they arise from
higher
dimensional operators. However in this case there is a very large
number of
operators that could be written down, and the restriction that just
one or two
of these operators contribute to the masses involves some strong
assumptions.

We are unable to completely avoid this dilemma: to obtain quark and
lepton mass
predictions from grand unified theories, assumptions about the
underlying flavor
structure of the theory must be made. Of all the flavor predictions
of GUTs
those pertaining to the heaviest generation are most direct and
subject to the
fewest assumptions. In this paper we pursue a scheme for predicting
the top mass which is unique in its simplicity. We attempt as
complete and accurate analysis of
this prediction as possible: the aim is to predict the top quark
mass to within
a few GeV.

There are many approaches in the literature which result in
predictions for the top mass. We would like to emphasize the
differences between two such approaches and a third one which we
shall take. The first is the infrared fixed point behaviour of the
renormalization group (RG)
equation for the top quark Yukawa coupling \cite{fixedpt}. This is
an argument
that certain values for the top mass are more probable than others
if all GUT scale Yukawa couplings are equally probable.
The second framework for predicting the top quark mass is that of
textures for
the generation structure of the Yukawa coupling matrices. The top
mass is given
in terms of lighter quark masses and entries of the
Kobayashi-Maskawa mixing
matrix \cite{gn,hrr}. This necessarily requires assumptions about
the masses of
the lighter generations. Finally, in the context of grand unified
theories, it
is sometimes possible to predict the top quark mass from a
consideration only of the heaviest generation, with the b and $\tau$
masses as inputs from experiment \cite{als,alsalt}. Consider the
Yukawa interactions of a supersymmetric grand unified theory which
lead to masses for the heaviest generation. There are three
observable masses: $m_t$, $m_b$ and $m_\tau$. Since the Higgs
doublet which leads to the top
mass is forced by supersymmetry to be different from the one which
gives
mass to b and $\tau$, these three masses necessarily depend on the
parameter
$\tan \beta$, the ratio of the vacuum expectation values (VEVs) of
these
two doublets.
If the grand unified theory has two or more independent Yukawa
parameters
contributing to the heaviest generation masses, then there will be
no
prediction when the heavy generation is considered in isolation. The
only
possibility for a prediction resulting from consideration of the
heaviest
generation alone is that the t, b and $\tau$ masses originate
predominantly
from a single Yukawa interaction. This immediately excludes
the grand unified gauge group SU(5) from consideration
\cite{sufive}. The
simplest supersymmetric \cite{nonsusy} grand unified gauge group
which allows relations between masses in the up and down sectors is
SO(10).

In this paper we study the top quark mass prediction which results
from the
following three assumptions:
\begin{enumerate}
\item[(I)] The masses of the third generation, $m_t$, $m_b$ and
$m_\tau$, originate
from renormalizable Yukawa couplings of the form
$\underline{16}_3\,{\cal O}\,\underline{16}_3$ in a supersymmetric
GUT with a gauge group containing (the conventional) SO(10).
\item[(II)] The evolution of the gauge and Yukawa couplings in the
effective theory beneath the SO(10) breaking scale
is described by the RG equations of the minimal supersymmetric
standard model (MSSM).
\item[(III)] The two Higgs doublets lie
predominantly in a single irreducible multiplet of SO(10).
\end{enumerate}
We find it highly significant that such simple and mild assumptions
are sufficient for predicting the top quark mass with an accuracy of
a few GeV, in terms of a few parameters of the MSSM (which could be
measured experimentally). In Secs.\ II--VII we in fact assume that
the two light
Higgs doublets lie completely in a single irreducible multiplet,
while in Sec.~IX we return to the effects of mixing with other
multiplets. We will very rapidly be led to the result that the
SO(10) multiplet containing
the Higgs doublets is (almost) necessarily the $\underline{10}_H$,
so that
the
relevant Yukawa interaction is
$\underline{16}_3\,\underline{10}_H\,\underline{16}_3$ (where
$\underline{16}_3$ is the third-generation matter supermultiplet).
The prediction of the top quark mass from this interaction was first
considered
by Ananthanarayan, Lazarides and Shafi \cite{als}. We find the
picture which emerges from such an interaction to be very
elegant. While the three Yukawa couplings $\lambda_{t,b,\tau}$ are
different at low energies, they evolve according to
RG equations to a common unified value at the
large mass
scale of the grand unified theory. The scenario is reminiscent of
the
evolution of the three gauge coupling constants to a common value at
the
unification scale. While the gauge coupling unification leads to a
prediction
for the weak mixing angle, the Yukawa coupling unification leads to
a prediction for the top quark mass. The top-bottom mass hierarchy
then originates in the Higgs sector, so another prediction is a
large ratio of Higgs VEVs. The weak mixing
angle prediction has
undergone several refinements as higher order corrections and a
variety of
threshold corrections have been considered. An aim of the present
paper
is to compute such corrections to the top mass prediction to a
similar level
of accuracy. In particular we study:
\begin{itemize}
\item the coupled two-loop RG equations for the
three gauge
couplings and the three Yukawa couplings. We give an analytic fit to
the
numerical results which is valid to better than $0.2\%$.
\item
some implications of generating the top-bottom mass hierarchy
through a large ratio of the VEVs of the two
light Higgs doublets.
This source
for up-down mass hierarchy is generic in models which unify all
three
Yukawa couplings of the third generation.
Since {\it a priori} there is no symmetry protecting the down-type
Higgs VEV and consequently the down-type fermion masses, large
radiative corrections to these masses typically arise and change the
top mass prediction considerably. Such corrections will be
suppressed if the squarks are much heavier than the higgsinos and
gauginos; indeed, we identify two symmetries which could then
protect the down-type VEV and masses. Whether such suppression is
favored in models with large $\tan\beta$ is a question for future
study \cite{newradsymm}, and whether it is the case in nature will
be determined by future experiments.
\item two consequences of these large corrections to the b quark
mass. For a certain range of MSSM parameters the $m_b/m_{\tau}$
prediction cannot be brought into agreement with experiment; for a
separate, smaller range, different GUT-scale boundary conditions
must be used.
\item the supersymmetric threshold corrections to the three gauge
couplings and the three Yukawa couplings. These are given for an
arbitrary spectrum of the superpartners of the minimal
supersymmetric standard model, ignoring only the electroweak
breaking effects in the spectrum. We find in particular that, when
the above symmetries hold approximately, raising any or all of the
superpartner masses increases $m_t$ for fixed $\alpha_3(m_Z)$.
\item threshold corrections at the grand unified mass scale. We show
that
such corrections to the gauge couplings do not significantly affect
the top mass prediction for a given $\alpha_3(m_Z)$.
We calculate the corrections to the Yukawa couplings from superheavy
splittings in the gauge \underline{45}, the $\underline{10}_H$
and the $\underline{16}_3$ multiplets (these corrections are not
very large), and give general expressions for further
possible superheavy threshold corrections.
\item the extent to which the predicted value of the top quark mass
depends on assumption (III). This assumption is, in our
opinion, the weakest part of the theoretical picture which underlies
the top
mass prediction (with the possible exception of the electroweak
symmetry breaking sector). Even if the third generation only couples
through a single
SO(10) invariant Yukawa interaction, the relation between this
coupling and the
values of $\lambda_t, \lambda_b$ and $\lambda_\tau$ renormalized at
the
unification scale may involve a set of mixing angles describing
which
two linear combinations of the doublets of the unified theory are
light.
 An understanding of these mixing angles is in principle related
to an understanding of why two doublets remain light (the
doublet-triplet
splitting problem). Our understanding of the resolution of this
problem
is at present not complete.
We are however greatly encouraged by the following two facts:
{\it i}) Due to the fixed point behavior of the RGE the prediction
is somewhat insensitive to these mixing angles in a large class of
models;
{\it ii}) in very simple SO(10) models which provide a partial
solution of the
doublet-triplet splitting problem, both Higgs
doublets are in fact contained within the same irreducible
representation (namely a
$\underline{10}_H$) \cite{babubarr,newradsymm}.
\item the extraction of the $b$ quark mass from
experiment, using updated experimental information and including the
dependence of the extracted value of
this mass on the QCD coupling $\alpha_3$. This is of importance to
us because the two crucial experimental inputs to the top mass
prediction are
$m_b$ and $\alpha_3$.
\end{itemize}
In Sec.~II we describe the basic framework for our calculation. A
discussion of the implications of large $\tan\beta$ is given in
Sec.~III, where we examine two potentially very large corrections to
the $b$ quark mass prediction; however, if the MSSM
parameters exhibit a certain hierarchy (for large $\tan\beta$)
such corrections may be suppressed.  In Sec.~IV
we give approximate analytic solutions to the two loop
renormalization group
equations, in the absence of threshold corrections. The extraction
of the b
quark mass from data is given in Sec.~V. Armed with this
experimental value of $m_b$, we return in Sec.~VI to the large
corrections to the predicted $m_b$, bounding these corrections (and
consequently the MSSM parameters) and investigating the possibility
of different GUT-scale initial conditions for the Yukawa couplings.
The remaining threshold corrections are
studied in Sec.~VII, while in Sec.~VIII the sensitivity of the
predicted top
quark mass to these threshold corrections is derived and discussed
in a general
way. In Sec.~IX we give the prediction for the top quark pole mass
and discuss its sensitivity to certain grand unified threshold
corrections. In Sec.~X we extend the discussion to include models
in which the doublets are not completely contained in a single
SO(10) irreducible multiplet, and to realistic models which include
the other generations of matter. Sec.~XI concludes.

\section{Framework}

The predicted value of the top mass depends on the renormalization
group equations (RGE), the boundary conditions at the GUT and
electroweak scales, and the threshold corrections at these two
scales. The one-loop RGE in the minimal supersymmetric standard
model (that is, with three generations, two Higgs doublets and no
right-handed neutrinos) are given by
\begin{mathletters}
\bea
16\pi^2 {d\ln\lt\over dt} &=& \sum_{\nu=t,b,\tau}
K_{t\nu}\lambda_\nu^2 + \sum_{i=1,2,3} L_{ti}g_i^2 \,
,\label{eqrgyukt} \\
16\pi^2 {d\ln\lb\over dt} &=& \sum_{\nu=t,b,\tau}
K_{b\nu}\lambda_\nu^2 + \sum_{i=1,2,3} L_{bi}g_i^2 \,
,\label{eqrgyukb} \\
16\pi^2 {d\ln\ltau\over dt} &=& \sum_{\nu=t,b,\tau}
K_{\tau\nu}\lambda_\nu^2 + \sum_{i=1,2,3} L_{\tau i}g_i^2 \,
,\label{eqrgyuktau}
\eea
\end{mathletters}
and
\begin{mathletters}
\bea
16\pi^2 {d\ln g_1\over dt} &=& b_1 g_1^2 \, ,\label{eqrggauone} \\
16\pi^2 {d\ln g_2\over dt} &=& b_2 g_2^2 \, ,\label{eqrggautwo} \\
16\pi^2 {d\ln g_3\over dt} &=& b_3 g_3^2 \, ,\label{eqrggauthree}
\eea
\end{mathletters}
where $t=\ln\mu$ and \cite{rge} $K_t = (6,1,0)$, $K_b = (1,6,1)$,
$K_\tau = (0,3,4)$, $L_t = (-{13\over15},-3,-{16\over3})$, $L_b =
(-{7\over15},-3,-{16\over3})$, $L_\tau = (-{9\over5},-3,0)$, $b_1 =
{33\over5}$, $b_2 = 1$, and $b_3 = -3$.
In our analysis, however, we employ two-loop evolution equations for
both the Yukawa and the gauge couplings. The gauge couplings are
related to experimental observables by $g_1^2 = {5\over3}
e^2/(1-\sin^2\theta_W)$, $g_2^2 = e^2/\sin^2\theta_W$, and $g_3^2 =
4\pi \alpha_3$, whereas the Yukawa couplings are related to the
running quark masses via $\lt = \sqrt{2} m_t/v_{U}$,
$\lambda_{b,\tau} = \sqrt{2} m_{b,\tau}/v_{D}$ and $v_{U}^2 +
v_{D}^2 = v^2 = (247\GeV)^2$. The VEVs of the two light Higgs
doublets $H_U$ and $H_D$ are denoted, respectively, by $v_U$ and
$v_D$, and their ratio is denoted by $v_{U}/v_{D}\equiv\tan\beta$ as
usual. The boundary conditions in the gauge sector at the
electroweak scale will be \cite{expdata,cpw} the $\msbar$ values
$\sin^2\theta_W = 0.2314$ (appropriate in anticipation of a heavy
top quark) and $4\pi/e^2 \equiv 1/\alpha_{\rm em} = 127.9$; these
are extracted from data using the 6-flavor standard model as the
effective theory at the scale $m_Z$. In the Yukawa sector we have
\cite{taumass} $m_\tau(m_\tau) = 1.777\GeV \Rightarrow m_\tau(m_Z) =
1.749\GeV$, and the range of $m_b$ values derived below. The
remaining uncertainties in $\sin^2\theta_W$ and $\alpha_{\rm em}$,
contribute negligibly to the prediction of $m_t$.

The tree-level initial conditions at the GUT scale $M_G$ are $g_1 =
g_2 = g_3 \equiv g_G$ for the gauge sector, but in the Yukawa sector
they depend on the source of the low-energy Yukawa couplings. In
SO(10) unification they typically arise from terms of the form
$\underline{16}_3\,{\cal{O}}\,\underline{16}_3$, where
$\underline{16}_3$ is the chiral supermultiplet for the third
generation and the Higgs multiplet $\cal{O}$ may be a
$\underline{10}_H$ or $\underline{126}_H$ of SO(10). (The
$\underline{120}_H$ is antisymmetric and therefore makes no
contribution.) To go beyond the SU(5) predictions and exploit the
larger SO(10) symmetries, we must {\it assume} that both light Higgs
doublets lie predominantly in a {\it single} SO(10) multiplet,
rather than being arbitrary mixtures of doublets in several
representations.
As stated above, our understanding of which doublets remain light,
and why they do so, is not complete. Thus to make progress we
simply assume the mixing is negligible. Note that if the mixing is
with other doublets which do not couple directly to the
$\underline{16}_3$, then the effect is only to split $\lt^G$ from
$\lambda_{b,\tau}^G$, and even a 30\% splitting of this sort will
have less than a 3\% effect on the low-energy top mass, as we show
below. Furthermore, a supersymmetric theory without mixing terms in
the Lagrangian is technically natural, due to the nonrenormalization
theorems.
Then either ${\cal{O}}\sim\underline{10}_H$, in which case the
initial conditions for the Yukawa couplings are
\beq
\lt = \lb = \ltau \equiv \lG\,,
\label{eqinitcond}
\eeq
or else ${\cal{O}}\sim\underline{126}_H$ from whence $3 \lt = 3 \lb
= \ltau \equiv \lG$. In the second case, a numerical investigation
shows that
the prediction of $m_b/m_\tau$ is too low unless
$\alpha_3(m_Z)>0.13$ and
$m_b(m_b)<4.0\GeV$, as well as $\lG > 4$. This last requirement,
however, when combined with the full SO(10) RGE, implies a Landau
pole in the Yukawa coupling less than 20\% above the unification
scale, which in general allows ${\cal O}(1)$ GUT-scale threshold
corrections
and makes any predictions impossible.
In fact, the only option for perturbative unification with
${\cal{O}}\sim\underline{126}_H$ is for very large electroweak-scale
threshold corrections to arise, which are of just the right
magnitude and sign to restore the agreement of $m_b/m_\tau$ with
experiment.  Only for a very limited range of $\lG$ values and MSSM
parameters can such a scenario be successful, as we show in Sec.~VI.
 For most of the parameter range, therefore, we need only
consider the consequences of
$\underline{16}_3\,\underline{10}_H\,\underline{16}_3$. In this
case, we must restrict $\lG<2$ to make sure the Landau pole is at
least a factor of 4 above the unification mass; for higher values of
$\lG$, we simply cannot make any reliable predictions, although
these values are by no means ruled out.

\section{Large $\protect\bbox{\tan\beta}$}

We discuss in this section some of the implications of the large
value of $\tan\beta\equiv v_U/v_D \simeq 50-60$ necessitated by the
boundary condition $\lt^G=\lb^G=\ltau^G$. In the standard model and
most of its extensions, the hierarchy $m_b\ll m_t$ (and similarly
for the $\tau$ mass) is imposed through $\lb\ll\lt$; $\lb$ is the
small parameter quantifying the breakdown of the chiral symmetry
which protects the bottom quark mass. In multi-Higgs models such as
the MSSM, there is also the option of explaining $m_b\ll m_t$ by
having the down-type Higgs acquire a much smaller VEV than that of
the up-type Higgs, $v_D\ll v_U$. We are {\it forced} to take this
second route by our GUT boundary condition, which implies
$\lb\sim\lt\sim1$.  We will assume
as usual that the electroweak symmetry is broken by an instability
of the scalar
potential $m_U^2 |H_U|^2 + m_D^2 |H_D|^2 + B\mu (H_U H_D +{\rm
h.c.}) + {1\over8} (g^2+g^{\prime 2}) (|H_U|^2-|H_D|^2)^2$  in the
$H_U$ direction. This
breaking is then communicated to $H_D$
by the soft SUSY-breaking term $\mu B H_U H_D$:
\beq
{-\mu B\over m_U^2+m_D^2} = \case{1}{2}\sin2\beta \simeq
{1\over\tan\beta}
=  {v_D\over v_U} \sim {1\over 50}.
\label{smallmu}
\eeq
So the magnitude of the hierarchy between the up and down sectors
is determined by $\mu B/(m_U^2+m_D^2)$, while its direction is set
by the direction in which the instability develops in the scalar
potential. The denominator is given by $m_U^2+m_D^2=m_A^2$, the
squared mass of
the pseudoscalar neutral Higgs boson; evidently, either the $\mu$ or
the $B$ parameter
or both must be much smaller than $m_A$ in order to generate the
top-bottom mass hierarchy. How, and indeed whether, the various
parameter values may arise will be discussed in a future paper
\cite{newradsymm}.

At tree level, an appropriate choice of parameters in the scalar
potential leads to $v_D\ll v_U$ and hence to $m_{b,\tau}\ll m_t$.
However, no symmetry has (yet) been imposed on the Lagrangian to
protect such a hierarchy, and therefore we expect large radiative
corrections. In fact corrections arise from the gluino- and
higgsino-exchange diagrams of Fig.~1 (the analogue of Fig.~1a, where
a
bino is exchanged, is suppressed by the small hypercharge coupling).
 Such corrections have been overlooked in past work
on large $\tan\beta$ scenarios, though they were discussed in the
context of radiative mass generation \cite{banks}.
Thus the MSSM prediction for $m_b$ becomes, after replacing $H_U
\rightarrow v_U/\sqrt{2}$,
\beq
m_b = \lb {v_D\over\sqrt{2}} + \delta m_b^{(\gl)} + \delta
m_b^{(\hi)}
\label{delmb}
\eeq
where
\beq
\delta m_b^{(\gl)} = {2\alpha_3\over3\pi} m_{\gl}\,\mu\lb
{v_U\over\sqrt{2}}
\,I(m_{\sb,+}^2,m_{\sb,-}^2,m_{\gl}^2)
\label{mbgl}
\eeq
and
\beq
\delta m_b^{(\hi)} = {\lt\lb\over16\pi^2} \mu\,A_t\lt
{v_U\over\sqrt{2}}
\,I(m_{\st,+}^2,m_{\st,-}^2,\mu^2) ,
\label{mbhi}
\eeq
$I$ is given in the appendix, $m_{\gl}$ is the gluino mass, $\mu$ is
the supersymmetric coupling of the two Higgs doublets, $A_t$ is the
trilinear soft SUSY-breaking coupling of the stop fields to the
up-type Higgs, and $m_{\sb,\pm}$
and $m_{\st,\pm}$ are the squark mass eigenstates propagating in the
loop. Numerically $m_{\gl}\simeq 3 m_{1/2}$ where $m_{1/2}$ is the
mass of either the gauginos at the GUT scale or of the wino at the
electroweak scale. To appreciate the significance of such
corrections, consider the limit in which the squarks have roughly
equal masses $m_0$, and $\mu$ or $m_{\gl}$ are either much less than
or equal to $m_0$. Then $I(m_0^2,m_0^2,0) = 1/m_0^2$ and
$I(m_0^2,m_0^2,m_0^2)= 1/(2 m_0^2)$, and we find:
\bea
m_b &=& \lb {v_D\over\sqrt{2}} \left[1 +
{\tan\beta\over16\pi^2}\left(
{8\over3} g_3^2 {m_{\gl}\mu\over(2)m_0^2} +
\lt^2 {\mu A_t\over(2)m_0^2} \right)\right]\cr
&\simeq&\lb {v_D\over\sqrt{2}} \left[1 +
0.35 \left(4\, {m_{\gl}\mu\over(2)m_0^2} +
{\mu A_t\over(2)m_0^2} \right)\right]
\label{fullmb}
\eea
(Note that in the second line above, we have approximated
$\tan\beta\simeq 50$, which is inaccurate if $\delta m_b$ becomes
large and lowers the top mass prediction significantly; we will
return to this point once we have extracted the experimental bounds
on $m_b$.) We see that the radiative corrections may in general be
comparable to the tree-level mass; when we equate this prediction to
the experimental value of $m_b$ to extract $\lb$, we would find
${\cal O}(1)$ corrections and hence large changes in $\lG$ and in
the prediction for $m_t$. Our final $m_t$ prediction would then be
very sensitive to the exact values of the squark, higgsino and
gaugino masses (and perhaps $A_t$ as well)---far more sensitive than
expected from ordinary threshold corrections. On the other hand, the
squarks may turn out to be relatively heavy, namely $m_0^2 \gg \mu
m_{\gl},\,\mu A_t$, in which case these corrections would be
suppressed. For example, if $m_0\simeq 1\TeV$ but $\mu\simeq
m_{\gl}\simeq A_t \simeq 200\GeV$ then $m_b$ changes by only $\sim
6\%$, which in turn corrects the top mass prediction by $\sim 4\%$.
Of course, the sign of this correction is determined by the sign of
the parameters which enter $\delta m_b$.

\begin{figure}
\leavevmode
\epsfxsize=14cm \epsfbox[-50 450 640 550]{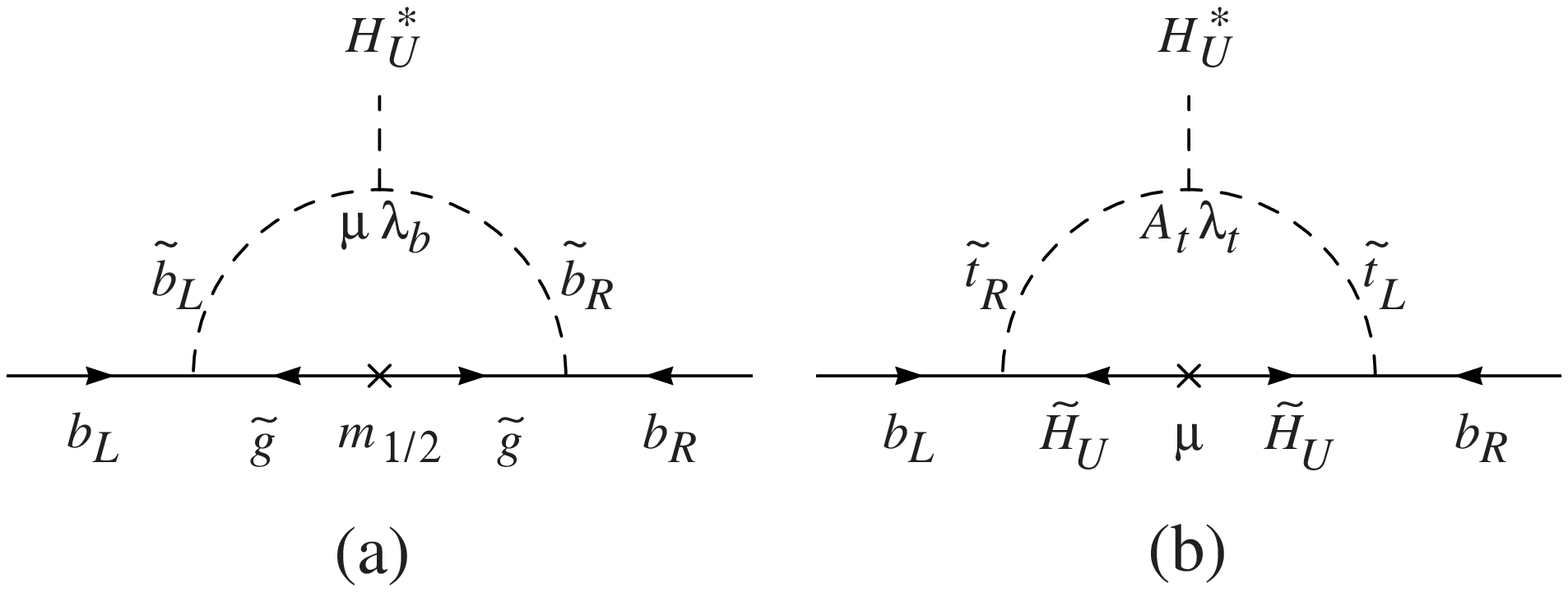}
\begin{quote}
{\small
FIG.~1. The leading (finite) 1-loop MSSM contributions to the $b$
quark mass.}
\end{quote}
\end{figure}

We should mention at this point that there is a diagram in
the 2-Higgs standard model analogous to the higgsino-exchange
diagram of Fig.~1, in which the stop propagator is replaced with a
top propagator, the higgsino with the Higgs, and the couplings
are replaced by $A_t\lambda_t \rightarrow \lambda_t$ and $\mu
\rightarrow \mu
B$. For this diagram the large $\tan\beta$ enhancement gained by
coupling the $b$ to $H_U^*$ is manifestly and exactly cancelled by
the $\mu B/m_A^2$ factor from the propagators, independent of any
symmetries. This threshold contribution is included in the function
$f_R$ defined in Sec.~V.

We have seen that when $m_0^2 \gg \mu m_{\gl},\,\mu A_t$ the large
radiative corrections are suppressed. We also know that for large
$\tan\beta$,
$m_A^2 \gg \mu B$. These hierarchies can be related by imposing two
approximate symmetries. The symmetry which sets $\mu$ to zero is a
Peccei-Quinn (\PQ)
symmetry under which the superfield $H_D$ and the SU(2)-singlet
bottom antiquark superfield $b^c$ have equal and opposite charges
while all other fields are invariant. It is explicitly broken only
by the soft SUSY-breaking term $\mu B H_U H_D$ and by the term $\mu
H_U H_D$ in the superpotential, so when treated as a spurion $\mu$
should be assigned a \PQ\ charge opposite to that of $H_D$. The
SUSY-breaking term contributes at tree level to the VEV of $H_D$,
while the supersymmetric one enters into (finite) loop diagrams
which correct the $b$ mass. We will quantify the degree to which
this symmetry is broken by the dimensionless parameter $\epspq
\equiv \mu/m_0$.

The symmetry which sets $B$ to zero must also set the gaugino mass
$m_{1/2}$ and the trilinear scalar coupling $A$
(and in particular $A_t$) to zero; note that $m_{1/2}$ and $A$
generate $B$ through the RG evolution. The desired symmetry
is in fact a continuous \R\  symmetry, and it is convenient to
choose an \R\  such that the superpotential is invariant while the
soft SUSY-breaking terms (except for the common scalar mass) is not.
Furthermore, we choose an \R\  under which the scalar $H_U$ is
invariant but the scalar $H_D$ and the $b$ quark mass operator $Q
b^c$ are not.
We will assign the superspace coordinate $\theta$ a charge of $+1$,
the superfield $H_U$ a charge 0, and the quark doublet superfield
$Q$ a charge 0; then the superfield $H_D$ carries a charge of $+2$,
the superfield $b^c$ carries a charge of 0, and the superfield $t^c$
carries a charge of $+2$. Thus the left-handed quarks have charge
$-1$, the SU(2)-singlet $b^c$ antiquark has charge $-1$ and the
SU(2)-singlet $t^c$ antiquark has charge $+1$. In a spurion
analysis, the soft SUSY-breaking parameters $m_{1/2}$, $A$, and $B$
carry a charge of $-2$. Once again, we define a dimensionless
symmetry-breaking measure $\epsr \equiv B/m_0 \lesssim A/m_0 \sim
m_{1/2}/m_0$.

When the \PQ\  or \R\  symmetries are approximately valid, we find
\beq
\delta m_b^{(\gl)} \sim {2\alpha_3\over3\pi} \left(\epspq  \epsr
\tan\beta\right) \left({m_{\gl}\over B}\right) m_b
= {2\alpha_3\over3\pi} \left({m_{\gl}\over B}\right)
\left({m_A^2\over m_0^2}\right) m_b
\label{mbglappr}
\eeq
and
\beq
\delta m_b^{(\hi)} \sim {\lt\lb\over16\pi^2} \left(\epspq  \epsr
\tan\beta\right) \left({A_t\over B}\right)
\left({\lt\over\lb}\right) m_b = {\lt^2\over16\pi^2} \left({A_t\over
B}\right) \left({m_A^2\over m_0^2}\right) m_b.
\label{mbhiappr}
\eeq
As expected, the large $\tan\beta$ enhancement was cancelled by the
$\epspq \epsr$ factor, provided $m_{\gl}\sim A_t\sim B$ and $m_A^2
\sim m_0^2$.

If either of these symmetries is to hold even approximately, there
must be a certain hierarchy in the supersymmetric spectrum. Notice,
however, that $\mu$ and $m_{1/2}$ are bounded below by LEP data
\cite{mubound}, at least for large $\tan\beta$:
numerically, they read roughly $(|\mu|-m_Z/2) (|m_{1/2}|-m_Z/2) >
m_Z/2$, so we expect both $\mu$ and $m_{1/2}$ to be at least as
large as $m_Z$. Thus the \PQ\  and \R\  symmetries can be meaningful
only if the scalar superpartners are significantly more massive than
the $Z$. This typically implies a degree of fine-tuning to get the
proper $Z$ mass, so it
remains to be seen \cite{newradsymm} whether it is more natural to
expect
$m_0 \gg m_Z \sim \mu \sim m_{1/2} \sim A_t \sim B$ and fine-tune
the $Z$ mass, or to expect $m_0 \sim m_Z \sim \mu \sim m_{1/2} \sim
A_t \gg B$ and fine-tune $B$. Perhaps electroweak symmetry breaking
does not arise from the usual running of the parameters in the Higgs
sector, but rather from some other mechanism (which does not
significantly alter the RG evolution of the gauge and Yukawa
couplings). Ultimately, experiment will decide what if any hierarchy
exists in these parameters. Some experimental information already
exists:
attaching a photon in all possible ways to the diagrams of Fig.~1
and inserting a flavor-changing vertex leads to an amplitude for
$b\rightarrow s\gamma$ which again is proportional to $\epspq
\epsr\tan\beta$. To reconcile this amplitude with the CLEO data on
$\Gamma(b\rightarrow s\gamma)$, we must either impose the \PQ\  and
\R\  symmetries or raise all the superpartner masses (since the
operator for
$b\rightarrow s\gamma$ is of dimension higher than 4). We will leave
these
constraints to future work \cite{newradsymm}.
(Analogous considerations can
be extended to other observables such as $B^0\overline{B}^0$,
$D^0\overline{D}^0$ and $K^0\overline{K}^0$ mixing, the electric
dipole moment of the neutron and proton decay.) For now, we can only
make definite
predictions of the top mass in the case $m_0^2 \gg \mu m_{\gl},\,\mu
A_t$.

\section{Running and Matching}

If we temporarily ignore all threshold corrections
(the ``unperturbed scenario''), the solution of the RG equations
proceeds schematically as follows. By requiring that the two gauge
couplings $g_1$ and $g_2$ meet, we solve
Eqs.~(\ref{eqrggauone},\ref{eqrggautwo}) to obtain $M_G$ and $g_G$.
(At this stage there is only a weak dependence on the Yukawa
couplings in the 2-loop RGE, and we may use representative values
for them.) Then by running back down with Eq.~(\ref{eqrggauthree})
we predict $g_3(m_Z)$ and hence $\alpha_3(m_Z)$. Next, we solve the
two equations (\ref{eqrgyukb},\ref{eqrgyuktau}) for the two unknowns
$\lG$ and $\tan\beta$, using $M_G$ and $g_G$ as well as $m_b(m_Z)$
and $m_\tau(m_Z)$ and the initial condition (\ref{eqinitcond}). In
practice, this step is simplified because $\tan\beta$ will always be
$\sim m_t(m_Z)/m_b(m_Z) \sim 50-60$, so we may set $\sin\beta$ to
unity from now on. We are then left with a single equation, namely
$m_b(m_Z)/m_\tau(m_Z) \equiv R = \lb(m_Z)/\ltau(m_Z)$, which we
solve for the single unknown $\lG$. Finally, we use $\lG$ to run
down with Eq.~(\ref{eqrgyukt}) and obtain $\lt(m_Z)$, which is then
used to determine the top quark mass.

More precisely, we will adopt the following procedure to correctly
incorporate 2-loop RG evolution with 1-loop matching conditions. We
choose to match the MSSM with the broken-electroweak standard model
using as an intermediate step the 2-Higgs standard model (2HSM).
This two-step procedure has two advantages: the presentation is
clearer, and the various matching contributions are easy to isolate.
\begin{itemize}
\item We first treat the gauge sector: we match the
experimentally-determined gauge couplings $g_i(m_Z)$ of the standard
model, which are essentially those in the 2HSM, to the couplings of
the MSSM by integrating in the superpartners. The conversion from
the $\msbar$ to the $\drbar$ scheme is numerically insignificant. We
then use 2-loop MSSM RGEs to run from $m_Z$ to some GUT scale
$\mu_G$ which we will {\it fix} to be some convenient value near
$10^{16}\GeV$; in this running we employ approximate values of the
Yukawa couplings. We thus calculate the gauge coupling boundary
values at this GUT scale.
\item Starting with these gauge couplings and with a given set of
Yukawa boundary conditions (collectively denoted by $\lG$ for the
moment) at the GUT scale $\mu_G$, we evolve the gauge and Yukawa
couplings with 2-loop MSSM RGEs in the $\drbar$ scheme to an {\it
arbitrary} electroweak scale $\mu_Z$ and obtain $\lambda_{t,b,\tau}
\equiv \lambda_{t,b,\tau}^{\rm MSSM,\drbar}(\mu_Z;\lG)$. These are
then matched to the 2-Higgs standard model, in which the
superpartners have been integrated out, to yield
$\lambda_{t,b,\tau}^{\rm 2HSM,\drbar}(\mu_Z;\lG,\{m_a\}) =
\lambda_{t,b,\tau}^{\rm MSSM,\drbar}(\mu_Z;\lG) [1 +
k_{t,b,\tau}(\mu_Z;\{m_a\})]$, where $\{m_a\}$ are the superpartner
masses.
\item We also start with the running $\msbar$ value of the $b$ quark
mass $m_b^{\msbar}(4.1\GeV)$, and evolve it with 2-loop QCD running
[that is, in the no-Higgs low-energy standard model (0HSM) having
only strong and electromagnetic interactions] to obtain $m_b^{\rm
0HSM,\msbar}(\mu_Z) = m_b^{\msbar}(4.1\GeV)/\eta_b$. Similarly we
run up the $\tau$ mass to obtain $m_\tau^{\rm 0HSM,\msbar}(\mu_Z) =
m_\tau/\eta_\tau$. Their ratio is defined to be $R^{\rm
0HSM,\msbar}(\mu_Z)$, which may be translated into the $\drbar$
scheme:
$R^{\rm 0HSM,\drbar}(\mu_Z) \simeq R^{\rm 0HSM,\msbar}(\mu_Z) [1 -
\alpha_3/3\pi]$. To match the 0HSM to the 2HSM requires some
knowledge of the Higgs sector masses. At tree-level and in the limit
of large $\tan\beta$, these are all given \cite{largetanbhiggs} in
terms of $m_Z$, $m_W$, and the mass $m_A$ of the physical neutral
pseudoscalar Higgs: the (mostly up-type) scalar $m_h \simeq m_Z$,
the other (mostly down-type) scalar $m_H \simeq m_A$, and the
charged (also mostly down-type) scalars $m_{H^\pm} =
\sqrt{m_W^2+m_A^2}$. We then obtain $R \equiv R^{\rm
2HSM,\drbar}(\mu_Z;m_A) = R^{\rm 0HSM,\drbar}(\mu_Z) [1 +
f_R(\mu_Z;m_A)]$.
\item Next we demand  $R = \lambda_{b}^{\rm
2HSM,\drbar}(\mu_Z;\lG,\{m_a\})/\lambda_{\tau}^{\rm
2HSM,\drbar}(\mu_Z;\lG,\{m_a\})$ which we may solve for $\lG$.
\item Finally, we use the $\lambda_{t}^{\rm
2HSM,\drbar}(\mu_Z;\lG,\{m_a\})$ corresponding to this $\lG$ and
calculate the top pole mass $m_t^{\rm pole} =
(v/\sqrt{2})\lambda_{t}^{\rm 2HSM}(\mu_Z;\lG,\{m_a\}) [1 +
f_t(\mu_Z;m_A)]$, defined as the position of the pole in the 2HSM
with perturbative QCD. The function $f_t$ contains the well-known
contribution from perturbative QCD radiative corrections as well as
often-neglected contributions from Yukawa radiative corrections.
Note that the observable $m_t^{\rm pole}$ must be independent of
$\mu_Z$ to 1-loop order; we have indeed checked that our final
values do not change by more than a GeV or so when we vary $\mu_Z$
around the electroweak scale. To be specific, we will use the value
$\mu_Z=m_Z$ in all the explicit values we present below.
\end{itemize}

We will make the following approximations when appropriate. First,
we use the full 1-loop threshold expressions involving $\lt^2$,
$\lb^2$, $\ltau^2$, and $g_3^2$, with the following exception: when
integrating out the superpartners and matching to the 2HSM, we are
neglecting operators of dimension $>4$ which are suppressed by the
superpartner masses; when calculating the top (rather than bottom or
tau) mass, this amounts to neglecting finite terms of order
$m_t^2/m_{\rm superpartner}^2$, which is not valid in the case of a
light superpartner but will numerically be sufficiently accurate.
Second, corrections proportional to electroweak gauge couplings have
only been included in leading $\log (m_{\rm SUSY})$ approximation;
this means that we have neglected finite parts from SUSY loops and
all electroweak gauge-boson loops.  Third, we keep only the dominant
diagrams (namely the ones in Fig.~1) from the class of finite
diagrams proportional to $\epspq \epsr$, $\epspq ^2$ and $\epsr^2$;
the rest contribute at most $\sim 1\%$ effects, or much less if
$\epspq <1$ and $\epsr<1$. Fourth, when integrating out the Higgs
sector we neglect the effects of various $1/\tan\beta$ mixings.
Finally, in the numerical results we have grouped together and
assigned common average masses to the squarks $(m_{\sq})$, the
sleptons $(m_{\sl})$, and the higgsinos $(m_{\hi})$.

The evolution of the Yukawa and gauge couplings from $\mu_G$ to
$\mu_Z$ in the MSSM, even to 1-loop accuracy, cannot in general be
calculated analytically. Numerical solutions are straightforward,
and show that, in order to obtain the correct $m_b/m_\tau$ ratio,
the GUT-scale Yukawa couplings must be ${\cal O}(1)$, so the top
mass is typically predicted near its fixed point value
\cite{als,alsalt,cpw}. With this in mind, it will be useful and
illuminating to obtain simple, explicit approximate solutions by
fitting $\lambda_{t,b,\tau}(\mu_Z)$ numerically to a quadratic
polynomial in $1/\lG$. The constant term reflects the independence
of the low-energy Yukawa couplings on the initial conditions in the
large-$\lG$ limit, a consequence of the fixed point behavior. When
varying $\lG$ between 0.5 and 2, we find in the unperturbed scenario
\beq
\lambda_{t,b,\tau}(\mu_Z) \simeq A_{t,b,\tau}+
{B_{t,b,\tau}\over\lG}+{C_{t,b,\tau}\over\lG^2}
\label{eqfitzero}
\eeq
where $A_{t,b,\tau} = (1.099,1.014,0.673)$, $B_{t,b,\tau} =
(-0.045,-0.012,-0.107)$, and $C_{t,b,\tau} = (-0.019,-0.025,0.001)$;
these values correspond to $\mu_Z = 90\GeV$ and to $g_3^G = g_G$
(that is, no superheavy threshold corrections) which leads to
$\alpha_3(m_Z) = 0.125$. We have checked that the errors we make in
this fit are smaller than $0.2\%$ over the entire range
$0.5\leq\lG\leq2.0$.
We may then solve the quadratic equation $R = (A_b+
B_b/\lG+C_b/\lG^2)/(A_\tau+ B_\tau/\lG+C_\tau/\lG^2)$ to obtain
$\lG$ as a function of $R$, which in turn gives an explicit solution
for $\lt(\mu_Z)$ as a function $F(R)$ of the
experimentally-determined ratio $R$. The functions
(\ref{eqfitzero}), as well as the ratio $R =
\lb(\mu_Z)/\ltau(\mu_Z)$, are plotted versus $\lG$ in Fig.~2a, while
$F(R)$ is shown in Fig.~3a.

\begin{figure}
\leavevmode
\epsfxsize=17cm \epsfbox[40 235 640 480]{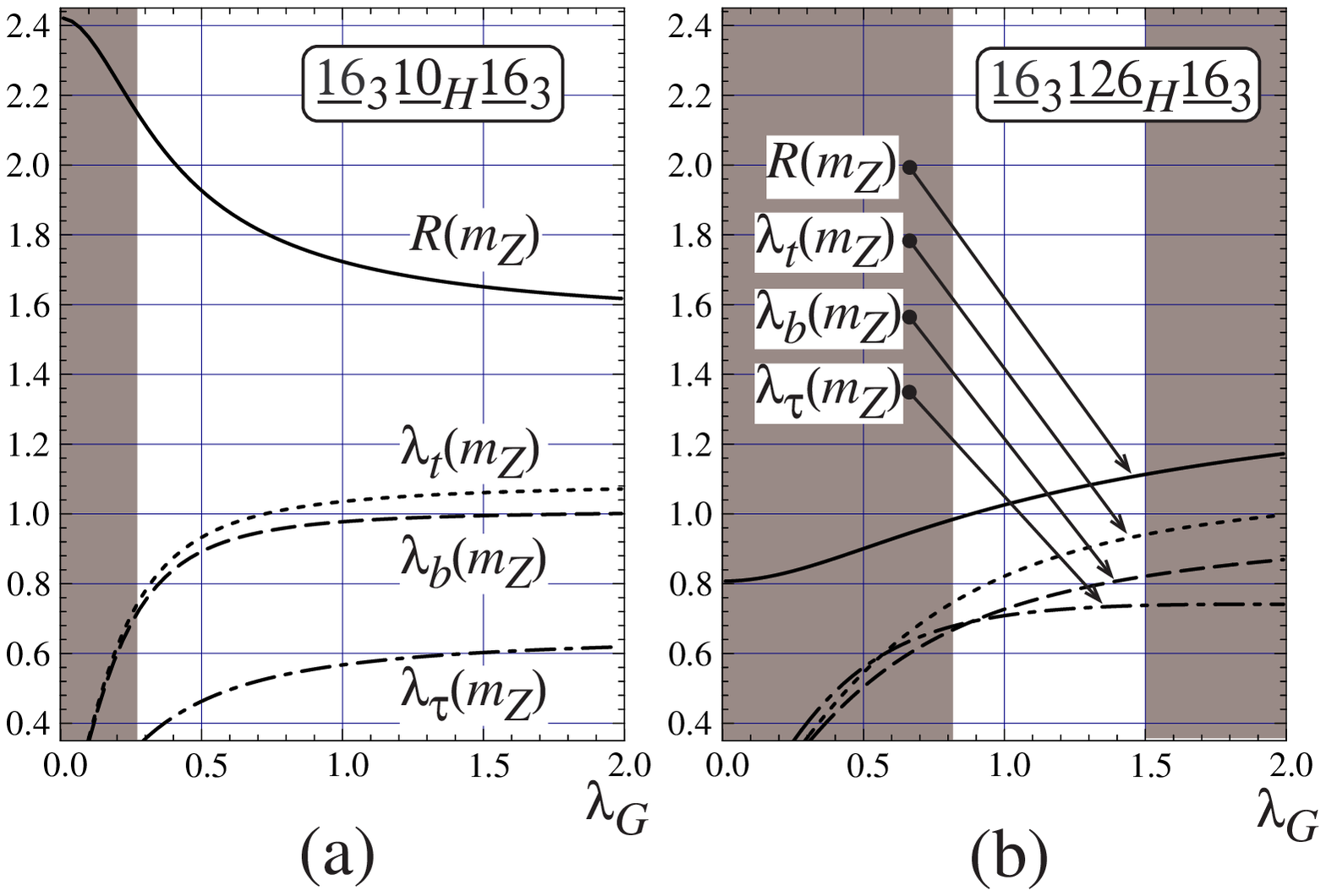}
\begin{quote}
{\small
FIG.~2. The dependence of the low-energy values
$\lambda_{t,b,\tau}(m_t)$ and of the ratio $R = \lb(m_Z)/\ltau(m_Z)$
on the initial condition $\lambda_{t,b,\tau}^G=\lG$ without any
threshold corrections is shown assuming the mass is generated
through (a) $\underline{16}_3\,\underline{10}_H\,\underline{16}_3$
or (b) $\underline{16}_3\,\underline{126}_H\,\underline{16}_3$. In
(a) the top mass exceeds 130 GeV only within the shaded region. In
(b) the coupling is perturbative only below $\lG \simeq 1.5$ and
$m_t>130\GeV$ only above $\lG \simeq 0.82$; thus the allowed region
is again the unshaded one.}
\end{quote}
\end{figure}

\begin{figure}
\leavevmode
\epsfxsize=17cm \epsfbox[40 240 640 520]{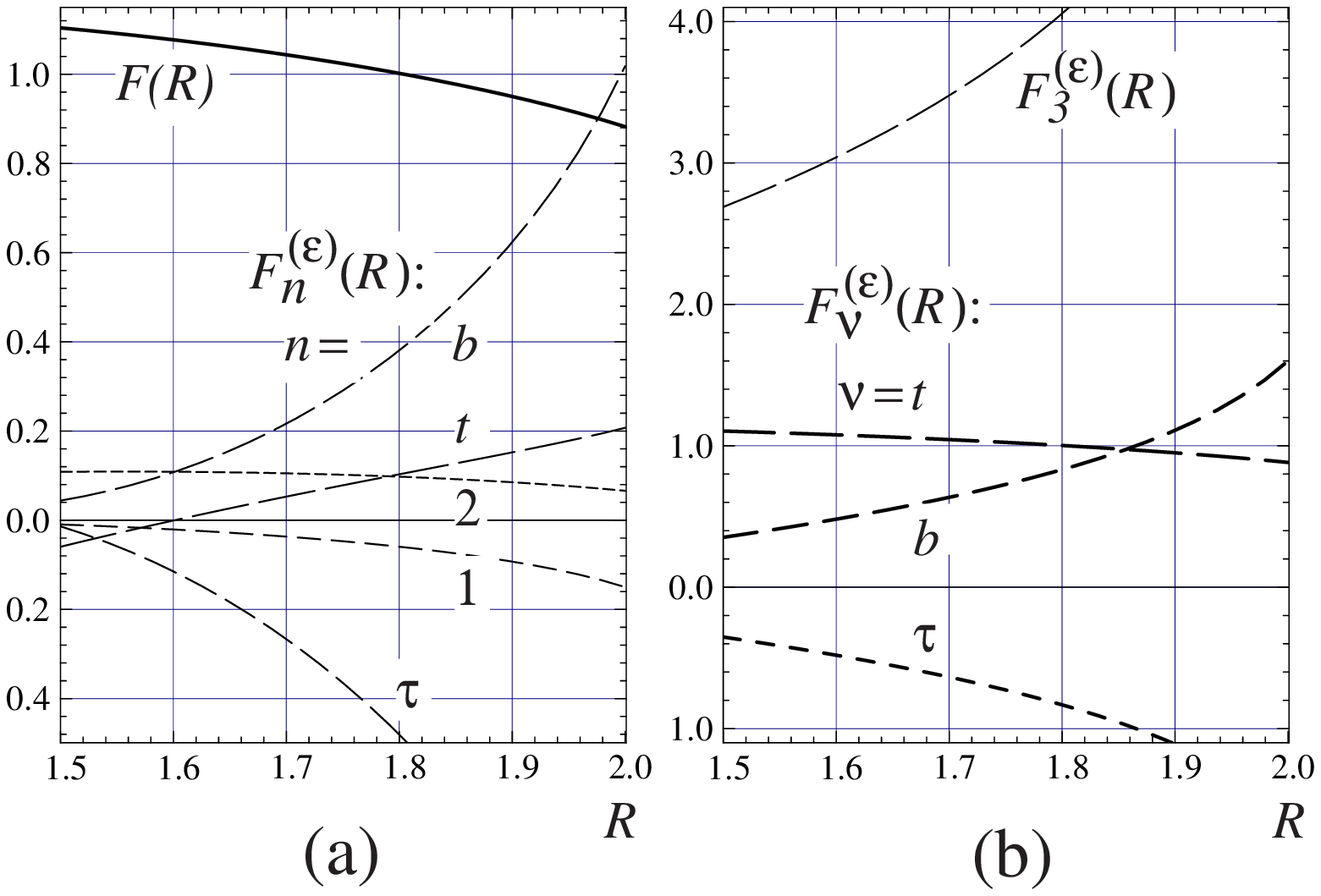}
\begin{quote}
{\small
FIG.~3.The induced dependence $\lt(m_Z)=F(R)$ derived from Fig.~2,
as
well as the sensitivity functions
$\{F_n^{(\epsilon)}(R)\}$ and $\{F_\nu^{(k)}(R)\}$.}
\end{quote}
\end{figure}

\section{The Mass of the Bottom Quark}

What is the experimental value of $R$? Using updated values from the
Particle Data Group compilation \cite{pdg}, we have reanalyzed the
well-known extraction of the b mass from $e^+e^-$ collisions via QCD
sum rules \cite{sumrules}.
The idea is to use a dispersion relation to relate the experimental
spectral distribution of $e^+e^-\rightarrow b\overline{b}$ to the
expectation value of the product of two vector $b$-quark currents.
This product is then rewritten as an operator product expansion;
perturbative QCD is used to calculate the coefficients of the
identity and other operators in the expansion, while the
nonperturbative information is assumed to be contained in the
condensates of these other operators. For the $b$ system, one
expects the nonperturbative terms to be at most ${\cal
O}(\langle\overline{\psi}\psi\rangle/m_b^3\sim\Lambda_{\rm
QCD}^3/m_b^3)$ and therefore negligible. The remaining calculation
is purely perturbative QCD, so the uncertainty in $m_b$ will be
dominated by our ignorance of the ${\cal O}(\alpha_3^2)$ terms in
the calculation of the coefficient of the identity operator. This
coefficient can be calculated reliably only for highly off-shell
momenta, for instance $q^2 \ll m_b^2$. Expanding the coefficient in
powers of $q^2/m_b^2$ results in a relation between the moments of
the spectral distribution and derivatives of the coefficient at $q^2
= 0$:
\bea
{\cal M}_n^{\rm expt} & = & {27\over4\pi\alpha_{\rm em}^2}
\left[\sum_V {\Gamma_V\over M_V^{2n+1}} + {\alpha_{\rm em}^2\over
27\pi} \left(1+{\alpha_3\over\pi}\right) {1\over n E_T^{2n}}\right]
\\
{\cal M}_n^{\rm theor}& = & {{\cal M}_{0n}\over m_{b,E}^{2n}}
\left(1 + A_n \alpha_3 + {\cal O}(\alpha_3^2)\right)
\label{eqsumrule}
\eea
where the sum approximating the spectral integral is over the
various resonances $V$ characterized by a mass $M_V$ and a width
$\Gamma_V$, $E_T \simeq 10.56\GeV$ is the estimated threshold energy
where the continuum $b\overline{b}$ production begins, and ${\cal
M}_{0n}$ and $A_n$ are numerical constants given in reference
\cite{sumrules}.

The mass is extracted from the ratios of the first few successive
moments,
\beq
r_n \equiv {{\cal M}_n\over {\cal M}_{n-1}} \simeq {r_{0,n}\over
m_{b,E}^2} \left(1 + a_n \alpha_3 + \kappa \alpha_3^2\right)
\label{eqratios}
\eeq
where $r_{0,2} = -0.00452$, $r_{0,3} = -0.00462$, $a_1 = -0.0286$,
$a_2 = -0.197$, and $\kappa$ has not yet been calculated but is
expected to be at most ${\cal O}(1)$ for the first few moments. We
will make a very rough estimate of our uncertainty by allowing
$\kappa$ to vary between $-2$ and $2$.
The strong coupling $\alpha_3$ in these expressions must be run down
from its given value at $m_Z$. The parameter $m_{b,E}$ is the
Euclidean pole mass of Georgi and Politzer \cite{gp}, which is
related to the running $\msbar$ mass we need via
\beq
m_{b,E} = m_b(\mu) \left[1 + {\alpha_3\over\pi} \left({4\over3} - 2
\ln 2 - 2 \ln {m_{b,E}\over\mu}\right)\right].
\label{eqmbrel}
\eeq
Using the first three moments, we obtain $3.93\GeV < m_b(4.1\GeV) <
4.36\GeV$ if $\alpha_3(m_Z) = 0.11$, and $3.86\GeV < m_b(4.1\GeV) <
4.42\GeV$ if $\alpha_3(m_Z) = 0.12$. The central values of
$m_b(4.1\GeV)$ extracted from the next few moments fall well within
these ranges, providing some confidence that our error bars are not
too small. Thus we estimate
\beq
m_b(m_b) = \left\{ \begin{array}{ll}
                              4.15\GeV\pm0.22 \GeV, &
\,\alpha_3(m_Z) = 0.11\\
                              4.14\GeV\pm0.28 \GeV, &
\,\alpha_3(m_Z) = 0.12
                                 \end{array}
                           \right. .
\label{eqmbpred}
\eeq
The central values we extract for $m_b$ are not very sensitive to
$\alpha_3$ since the $a_n\alpha_3\ll1$; they are somewhat lower than
in the older analyses mainly because the more precise experimental
value we use for the electronic partial width of the
$\Upsilon(9460)$ is higher than the older value. Our error bars in
$m_b$ are larger than those of Gasser and Leutwyler \cite{gl}
because of the different ways we estimate the error from the
${\cal O}(\alpha_3^2)$ terms.

The 2-loop QCD evolution between $4.1\GeV$ and $\mu_Z$ reduces the
mass of the $b$ by a factor $\eta_b \simeq 1.437 + 0.075
[\alpha_3(m_Z)-0.115]/0.01$ for $\mu_Z = 90\GeV$; the corresponding
electromagnetic reduction factor for $m_\tau$ is $\eta_\tau\simeq
1.016$. The translation from $\msbar$ to $\drbar$ increases $m_b$ by
roughly half a percent and has virtually no effect on $m_\tau$.
Finally, we match the 0HSM model to the 2HSM by including the
radiative corrections of the Yukawa couplings via the function
$f_R(m_A)$,
\bea
f_R(m_A) &\equiv& R^{\rm 2HSM}/R^{\rm 0HSM} - 1
\nonumber\\
&\simeq& -0.014 \ln(m_A/m_Z)\quad{\rm when}\quad
\mu_Z = m_Z \quad{\rm and}\quad m_A > m_Z
\label{fsubr}
\eea
and arrive at the final value of $R$:
\bea
R &=&{m_b^{\msbar}(4.1\GeV)\over m_\tau} {\eta_\tau \over \eta_b}
\left[1-{\alpha_3(\mu_Z)\over 3\pi}\right] \left[1 + f_R(m_A)\right]
\label{eqRpred}\\
& = & \left\{ \begin{array}{lll}
     1.67\pm0.09, & \,\alpha_3(m_Z) = 0.11, & \,m_A = 90\GeV\\
     1.58\pm0.11, & \,\alpha_3(m_Z) = 0.12, & \,m_A = 90\GeV\\
     1.62\pm0.09, & \,\alpha_3(m_Z) = 0.11, & \,m_A = 1000\GeV\\
     1.53\pm0.11, & \,\alpha_3(m_Z) = 0.12, & \,m_A = 1000\GeV .
                                 \end{array}
                           \right.
\nonumber
\eea
The exact expression for $f_R$ may be found in the appendix.
Notice that a heavy second Higgs decreases the apparent experimental
value of $R$, or more intuitively increases the GUT prediction of
$R$---hence to agree with the experimental value, we need to lower
the prediction of $R$, which entails raising $\lG$ and with it
$m_t$. The lightest top masses result from a light Higgs sector.

\section{Implications of the Large Corrections}

With the experimental value of $m_b$ in hand, we can return to the
threshold corrections of Eq.~(\ref{fullmb}) to bound the allowed
range of MSSM parameters and to possibly allow different initial
conditions for the Yukawa couplings at the GUT scale.

First, let us see how large can
the corrections to $m_b$ become before $m_b$ as predicted from
Yukawa unification disagrees with the experimental range
given above. We focus on the (usually dominant) gluino contribution;
similar considerations will bound the higgsino diagram. It is
convenient to first divide Eq.~(\ref{fullmb}) by $m_\tau$, and
substitute $\tan\beta = (m_t/m_\tau) (\ltau/\lt) = (m_t/m_\tau)
(\lb/\lt) R_{\rm MSSM}^{-1} \simeq
0.95 R_{\rm MSSM}^{-1} (m_t/m_\tau)$ and
$I(m_{\sb,+}^2,m_{\sb,-}^2,m_{\gl}^2) \equiv 1/m_{\rm eff}^2$, to
obtain:
\bea
R_{\rm exp} = {m_b\over m_\tau} &=& R_{\rm MSSM}
+ \left(0.95 {8\over3}{\alpha_3\over4\pi}\right)
{m_t(R_{\rm MSSM})\over m_\tau} \left({m_{\gl}\mu\over m_{\rm
eff}^2}\right) \cr
&\simeq& R_{\rm MSSM} + {m_t(R_{\rm MSSM})\over 73\GeV}
\left({m_{\gl}\mu\over m_{\rm eff}^2}\right).
\label{Rshift}
\eea
Note that $m_t$ depends on $\lG$ which is in turn determined by
$R_{\rm MSSM}$. The $R_{\rm exp}$ on the left-hand side is the
experimental value $R$ extracted above, while $R_{\rm MSSM}$ on the
right-hand side denotes the value of $\lb/\ltau$ obtained by running
down in the MSSM from the GUT scale. Also, to first approximation
$m_t(R_{\rm MSSM}) \simeq (174\GeV)\, \lt(R_{\rm MSSM})$. As we will
establish from Eq.~(\ref{eqrgRappr}) below, and as illustrated in
Fig.~2a, $R_{\rm MSSM}$ is
bounded from below by roughly 1.6, corresponding to $\lG \rightarrow
\infty$; and from above by roughly 2.4, corresponding to $\lG
\rightarrow 0$.  Since the latter limit also corresponds to $m_t
\rightarrow 0$, it can be improved by enforcing the experimental
bound $m_t > 130\GeV$, so we conclude that $1.6 < R_{\rm MSSM} <
2.15$. We use these bounds and those of Eq.~(\ref{eqRpred}) to set
limits on $\left({m_{\gl}\mu/ m_{\rm eff}^2}\right)$: if this
quantity is positive, an upper bound results from taking the
smallest $R_{\rm MSSM}$ possible and the largest $R_{\rm exp}$
allowed, in which case $m_t$ is fixed at its maximal value; if that
quantity is negative, the largest $R_{\rm MSSM}$ and smallest
$R_{\rm exp}$ are needed, in which case $m_t \simeq 130\GeV$. In
this way we find the surprisingly stringent limits
\beq
-0.37\, \alt\, {m_{\gl}\mu\over m_{\rm eff}^2}\, \alt\, 0.08
\label{muglimits}
\eeq
which are obviously phenomenologically interesting signatures, but
are also relevant to the electroweak symmetry-breaking sector of
this model \cite{newradsymm}.

Second, recall that if we impose the GUT-scale initial condition $3
\lt = 3 \lb = \ltau \equiv \lG$ corresponding to mass generation
through a $\underline{126}_H$, then after evolving down to the
electroweak scale the prediction for $R$ is too small. Large $\delta
m_b$ corrections can restore agreement with experiment. Fig.~2b
shows $R$ as well as $\lt$, $\lb$ and $\ltau$ at the electroweak
scale as functions of $\lG$. In the presence of the
$\underline{126}_H$, the coupling $\lG$ must be kept below $\sim
1.5$ (rather than $\sim2$ for the $\underline{10}_H$) to raise the
Landau pole by a factor of 4 above the unification mass. Also, since
$m_t \sim (174\GeV)\lt$, the lower bound of $m_t > 130\GeV$ implies
$\lt > 0.75$ and hence from the figure $\lG > 0.82$. Within this
restricted range, $R_{\rm MSSM}$ varies between $1.0$ and $1.1$, so
to reconcile this with $R_{\rm exp}$ (now using $\lb/\lt \simeq
0.89$ and again focusing on the gluino diagram) requires
\beq
0.24\, \alt\, {m_{\gl}\mu\over m_{\rm eff}^2}\, \alt\, 0.42\,.
\label{mugrange}
\eeq
We learn that {\it if} the gluino, higgsino and squark mass
parameters satisfy $m_{\gl}\mu/ m_{\rm eff}^2 = 0.33 \pm 0.09$ and
{\it if} $\lG \simeq 1.1 \pm 0.3$ then the mass may originate
perturbatively from the coupling
$\underline{16}_3\,\underline{126}_H\,\underline{16}_3$. Notice that
in such a scenario, even if $m_{\gl}\mu/ m_{\rm eff}^2$ is known
precisely, then the uncertainty in $R_{\rm exp}$ is large enough
that the top mass prediction will usually be very imprecise.

Thus for Yukawa unification in the MSSM it is useful to distinguish
four
regions of parameter space:
\begin{enumerate}
\item If $|m_{\gl}\mu/ m_{\rm eff}^2 |\ll 1$ then the $\delta m_b$
corrections may be ignored, the mass must originate from a
$\underline{16}_3\,\underline{10}_H\,\underline{16}_3$ interaction,
and we can predict $m_t$ with little further dependence on the MSSM
parameters, as shown below;
\item If $-0.37\, \alt\, {m_{\gl}\mu/ m_{\rm eff}^2}\, \alt\, 0.08$
then the mass must still arise from a $\underline{10}_H$, but now
the prediction for $m_t$ depends very sensitively upon $m_{\gl}\mu/
m_{\rm eff}^2$ and can vary over the full  experimentally-allowed
range;
\item If $0.24\, \alt\, {m_{\gl}\mu/ m_{\rm eff}^2}\, \alt\, 0.42$
then the $\underline{126}_H$ must be used, with $\lG \simeq 1.1 \pm
0.3$. In this case the prediction, while imprecise, tends to be in
the lower half of the experimentally-allowed range;
\item If $m_{\gl}\mu/ m_{\rm eff}^2$ lies outside of these ranges
then perturbative Yukawa unification under our assumptions cannot be
reconciled with experiment.
\end{enumerate}

\section{Threshold Corrections}

Next, we investigate the deviations in the top mass prediction
induced by threshold corrections at the GUT and SUSY scales. For
convenience, we choose to {\em always} match the full SO(10) theory
with the MSSM at the scale $M_G$ of the unperturbed scenario, where
$g_1$ and $g_2$ met: we {\it define} from now on
$\mu_G=2.7\times10^{16}\GeV$. The top mass prediction as a function
of $R$ is completely determined once we know the initial conditions
$\{\lt^G,\lb^G,\ltau^G,g_1^G,g_2^G,g_3^G\}$ at the fixed scale
$\mu_G$, as well as the functions $k_{t,b,\tau}(\{m_a\})$ of the
superpartner masses which match between the MSSM (in which we run
with the RGE) and the 2-Higgs standard model (in which we calculate
the top pole mass). So we first calculate the perturbations
$\epsilon_{t,b,\tau,1,2,3}$ to the initial conditions and the
matching functions $k_{t,b,\tau}$ in terms of the various mass
thresholds and $\alpha_3(m_Z)$. Then we study the sensitivity of the
top mass prediction to these $\epsilon$'s and $k$'s. A linear
analysis of these perturbations is sufficient for our purposes.

Define for convenience
\begin{mathletters}
\bea
t_a &\equiv& \ln (m_a / m_Z), \qquad
T_\alpha \equiv \ln (M_\alpha / \mu_G),
\label{eqta}\\
t_{a,b} &\equiv& G_1(m_a^2/\mu_Z^2,m_b^2/\mu_Z^2)
\nonumber\\
&\simeq& \left\{ \begin{array}{lll}
     \ln(m_a/\mu_Z)-{1\over4}, & \, m_a\gg m_b ,\\
     \ln(m_{a=b}/\mu_Z), & \, m_a = m_b ,\\
     \ln(m_b/\mu_Z)-{3\over4}, & \,m_a\ll m_b ,
                                 \end{array}
                           \right.
\label{eqtb}\\
t'_{a,b,c} &\equiv& \ln{\max(m_a,m_b,m_c)\over\mu_Z},
\label{eqtc}
\eea
\end{mathletters}
where
$G_1$ is given in the appendix, $\{m_a\}$ are the masses of the
various superpartners and $\{M_\alpha\}$ are those of the superheavy
particles. The functions $t_a$ and $t_{a,b}$ yield the exact
threshold corrections and will be used in the dominant terms in
$k_{t,b,\tau}$ below. The function $t'_{a,b,c}$ only yields
threshold corrections in the leading $\log (m_{\rm SUSY})$
approximation and will be used in the subdominant terms.
The contributions of the superpartners to the gauge
$\beta$-functions will be denoted by $b_i^a$.
Superheavy threshold corrections can arise from couplings dressing
the $\underline{16}_3\,\underline{10}_H\,\underline{16}_3$ vertex
and having strength $g_G$ (the gauge \underline{45}), $\lambda_G$
(the $\underline{16}_3$ and $\underline{10}_H$), or some other
$\lambda'_A$. We write the corresponding contributions of any
superheavy particle with mass $M_\alpha$ to the Yukawa RGE by
$L_{\nu}^\alpha g_G^2$, $K_{\nu}^\alpha \lG^2$ and $\sum_A K_{\nu
A}^{\prime\alpha} \lambda_A^{\prime 2}$. We denote squarks,
sleptons, higgsinos, winos, binos and gluinos by $\sq$, $\sl$,
$\hi$, $\wi$, $\bi$, and $\gl$, and define $\alpha_G \equiv
g_G^2/4\pi$ and $y_x \equiv \lambda_x^2/4\pi$ for any $x$. We will
neglect all electroweak-breaking effects in the mass splittings, and
the few subdominant corrections mentioned below.  We expect all such
effects to alter $m_t$ by less than a GeV or so.

The gauge couplings at $\mu_G$ may be completely determined by their
low-energy values $\alpha_{1,2,3}(m_Z)$ and by the SUSY spectrum
simply by running them up from the Z mass to $\mu_G$.
When the superpartner masses are changed away from the 90 GeV value
of the unperturbed scenario, the couplings $g_1^G$ and $g_2^G$ at
$\mu_G$ deviate from the common value $g_G = 0.730$ of the
unperturbed scenario in an easily-calculable way. Similarly, $g_3^G$
deviates from this $g_G$ value when the masses of the colored
superpartners changes and also when we vary \cite{aszpred}
$\alpha_3(m_Z)$ away from $0.125$. While these changes to $g_i^G$
are calculated using the experimentally-accessible quantities
$\{m_a\}$ and $\alpha_3(m_Z)$, from a top-down viewpoint they should
be regarded as the net threshold corrections resulting from
integrating out the superheavy (or Planck-scale \cite{gravsm})
degrees of freedom in the SO(10) theory to arrive at the MSSM.
The initial conditions for the gauge couplings become
\begin{mathletters}
\bea
g_{1G} &\equiv& g_G (1+\epsilon_1) = g_G \left[1 -
{\alpha_G\over4\pi}
\left({11\over10} t_{\sq} + {9\over10} t_{\sl} + {2\over5}
t_{\sqrt{2}\hi}\right)
\right]\,, \label{eqeps1} \\
g_{2G} &\equiv& g_G (1+\epsilon_2) = g_G \left[1 -
{\alpha_G\over4\pi}
\left({3\over2} t_{\sq} + {1\over2} t_{\sl} + {4\over3}
t_{\sqrt{2}\wi}
 + {2\over3} t_{\sqrt{2}\hi}\right)\right]\,,
\label{eqeps2} \\
g_{3G} &\equiv& g_G (1+\epsilon_3) = g_G \left[1 -
{\alpha_G\over4\pi}
\left(2 t_{\sq} + 2 t_{\sqrt{2}\gl}\right) +
{1\over2} {\alpha_G\over\alpha_3(m_Z)}
\left(\delta\alpha_3\over\alpha_3\right) \right] \label{eqeps3}
\eea
\end{mathletters}
where  $\delta\alpha_3/\alpha_3 \equiv \left[\alpha_3(m_Z) -
0.125\right]/0.125$. (Note that the fermionic superpartners $\wi$,
$\hi$ and $\gl$ must be integrated out at $\sqrt{2}$ times their
mass.) The Yukawa couplings at $M_G$ differ from $\lG$ due to GUT
thresholds effects, and hence
\begin{mathletters}
\label{eqeps}
\bea
\lambda_{tG} &\equiv& \lG (1+\epsilon_t) = \lG \left[1 +
{1\over4\pi}\sum_\alpha
(K_t^\alpha y_G + L_t^\alpha \alpha_G + \sum_A K_{t
A}^{\prime\alpha} y'_A) T_\alpha \right]\,,
\label{eqepst} \\
\lambda_{bG} &\equiv& \lG (1+\epsilon_b) = \lG \left[1 +
{1\over4\pi}\sum_\alpha
(K_b^\alpha y_G + L_b^\alpha \alpha_G + \sum_A K_{b
A}^{\prime\alpha} y'_A) T_\alpha \right]\,,
\label{eqepsb} \\
\lambda_{\tau G} &\equiv& \lG (1+\epsilon_\tau) = \lG \left[1 +
{1\over4\pi}\sum_\alpha
(K_\tau^\alpha y_G + L_\tau^\alpha \alpha_G + \sum_A K_{\tau
A}^{\prime\alpha} y'_A) T_\alpha \right]\,.\label{eqepstau}
\eea
\end{mathletters}
Two remarks should be made at this point. First, the SUSY threshold
corrections actually enter Eqs.~(\ref{eqeps}) indirectly, even if
all superheavy particles are degenerate, since these corrections
generically change the scale at which $g_1$ and $g_2$ meet and
therefore change the predicted average mass of the superheavy
particles. The result is a shift in all the $T_\alpha$ that is
independent of $\alpha$. Second, any effect that contributes equally
to $\epsilon_t$, $\epsilon_b$ and $\epsilon_{\tau}$ makes no
contribution to our final result, since this only amounts to a
redefinition of $\lG$.

Finally, the superpartner masses induce threshold corrections to the
Yukawa couplings when they are integrated out of the MSSM to yield
the 2-Higgs standard model. We find
\begin{mathletters}
\bea
4\pi k_t &=& y_t \left({\textstyle{1\over2}}t_{\tr,\hi} +
t_{\tl,\hi}\right) +
{\textstyle{1\over2}} y_b t_{\br,\hi} + {\textstyle{4\over3}}
\alpha_3 \left(t_{\tl,\gl} + t_{\tr,\gl}\right) +
3\alpha_2 \left({\textstyle{1\over4}} t_{\tl,\wi} +
{\textstyle{1\over2}} t_{\hi,\wi} -
t'_{\wi,\tl,\hi}\right) \nonumber \\
&\phantom{=}& +
{\textstyle{3\over20}}\alpha_1 \left({\textstyle{1\over9}}
t_{\tl,\bi} + {\textstyle{16\over9}} t_{\tr,\bi} +
2 t_{\hi,\bi} - {\textstyle{16\over3}} t'_{\bi,\hi,\tr} +
{\textstyle{4\over3}}t'_{\bi,\hi,\tl}\right)\,, \label{eqkt} \\
4\pi k_b &=& y_b \left({\textstyle{1\over2}}t_{\br,\hi} +
t_{\bl,\hi}\right) +
{\textstyle{1\over2}} y_t t_{\tr,\hi} + {\textstyle{4\over3}}
\alpha_3 \left(t_{\bl,\gl} + t_{\br,\gl}\right) +
3\alpha_2 \left({\textstyle{1\over4}} t_{\bl,\wi} +
{\textstyle{1\over2}} t_{\hi,\wi} -
t'_{\wi,\bl,\hi}\right) \nonumber \\
&\phantom{=}& +
{\textstyle{3\over20}}\alpha_1 \left({\textstyle{1\over9}}
t_{\bl,\bi} + {\textstyle{4\over9}} t_{\br,\bi} +
2 t_{\hi,\bi} - {\textstyle{8\over3}} t'_{\bi,\hi,\br} -
{\textstyle{4\over3}}t'_{\bi,\hi,\bl}\right) +
4\pi k_b'\,,
\label{eqkb} \\
4\pi k_\tau &=& y_\tau \left({\textstyle{1\over2}}t_{\taur,\hi} +
t_{\taul,\hi}\right) +
3\alpha_2 \left({\textstyle{1\over4}} t_{\taul,\wi} +
{\textstyle{1\over2}} t_{\hi,\wi} -
t'_{\wi,\taul,\hi}\right) \nonumber \\
&\phantom{=}& +
{\textstyle{3\over20}}\alpha_1 \left(t_{\taul,\bi} + 4 t_{\taur,\bi}
+
2 t_{\hi,\bi} - 8 t'_{\bi,\hi,\taur} +
4 t'_{\bi,\hi,\taul}\right) \label{eqktau}
\eea
\end{mathletters}
where the $y_\nu$ and $\alpha_i$ are evaluated at $\mu_Z$.The
non-logarithmic threshold correction of Eq.~(\ref{fullmb}) must be
included if the squarks are not much heavier than $\mu$, $m_{\gl}$
and $A_t$:
\beq
k_b' = {\tan\beta\over4\pi}\left({8\over3}\alpha_3
{m_{\gl}\mu\over m_{\rm eff}^2} + {\lt^2\over4\pi}
{\mu A_t\over m_{\rm eff'}^2}\right)
\label{newk}
\eeq
where $m_{\rm eff}^2\equiv 1/I(m_{\sb,+}^2,m_{\sb,-}^2,m_{\gl}^2)$
and $m_{\rm eff'}^2\equiv 1/I(m_{\st,+}^2,m_{\st,-}^2,\mu^2)$.

Note
that the majority of the terms in $k_\nu$ are due to wavefunction
renormalization from scalar-fermion interactions and hence increase
with the superpartner masses; the eventual conclusion will be that
the smallest $m_t$ is predicted when all superpartners are as light
as possible (if $k_b'$ may be neglected, as discussed in Sec.~III).

The mass of the Higgs bosons (which are determined by $m_W$ and
$m_A$) enter the top prediction through the matching of $R$ between
the 0HSM and the 2HSM, and through the calculation of the top quark
pole mass in the 2HSM.
The former was included in the previous section as a correction to
$R$, while the latter is studied in Sec.~IX.

\section{Sensitivity Functions}

When the initial conditions are perturbed, so are the coefficients
of the fit in Eq.~(\ref{eqfitzero}). We expect the fit parameters to
vary linearly with the perturbations as long as these are
sufficiently small, and so we write
\beq
\lambda_\nu(\mu_Z) = \left[(A_\nu+{\displaystyle \sum_n} a_{\nu
n}\epsilon_n) + {B_\nu+{\displaystyle \sum_n} b_{\nu
n}\epsilon_n\over\lG} + {C_\nu+{\displaystyle \sum_n} c_{\nu
n}\epsilon_n\over\lG^2}\right] (1 + k_\nu)
\label{eqfit}
\eeq
where the sum ranges over $n=t,b,\tau,1,2,3$. The new fit
coefficients $\{a_{\nu n},b_{\nu n},c_{\nu n}\}$ must be computed
numerically. We then solve $R=\lb(m_t)/\ltau(m_t)$ for $\lG$ as
before, substitute back into Eq.~(\ref{eqfit}) and expand to first
order in $\epsilon_n$ and $k_\nu$ to find
\beq
\lt(\mu_Z) = F(R) + \sum_n F_n^{(\epsilon)}(R) \epsilon_n + \sum_\nu
F_\nu^{(k)}(R) k_\nu .
\label{eqtsol}
\eeq
$F(R)$ and the five ``sensitivity functions''
$F_{t,b,\tau,1,2}^{(\epsilon)}(R)$ are plotted in Fig.~3a while the
four sensitivity functions $F_3^{(\epsilon)}(R)$ and
$F_{t,b,\tau}^{(k)}(R)$ appear in Fig.~3b. We have checked that, for
the entire range $0.105\leq\alpha_3(m_Z)\leq0.13$ and $\{m_a\}\leq
3\TeV$, our approximation is off by at most $\sim 1\%$.

We can understand the behavior of $\lt(\mu_Z)$ shown in Figs.~2a and
3a,b as
follows.
\begin{itemize}
\item First, as is well known \cite{fixedpt}, $\lt$ and $\lb$ (and
to a lesser extent $\ltau$) are quite insensitive to $\lG$ for large
$\lG$, since they both tend towards a fixed point as
$\lG\rightarrow\infty$. The fixed-point behavior of $\lt$ is
manifested in the smallness of $F_t^{(\epsilon)}(R)$ --- changing
the initial $\lt^G$ by 10\% changes its final value $\lt(\mu_Z)$ by
at most 1\%. The sensitivity of $\lt(\mu_Z)$ to $\lt^G$ is even less
for small $R$, since small $R$ implies large $\lG$ and hence a
stronger fixed-point behavior for $\lt$.
\item The value at the fixed point is determined predominantly by
$\alpha_3$ at low energies, and hence $F_3^{(\epsilon)}(R)$ is
large; conversely, $\alpha_{1,2}$ are smaller at low energies and
hence have little influence, so $F_{1,2}^{(\epsilon)}(R)\ll1$.
\item Next, we establish the fixed-point nature of
$\lb(t)/\ltau(t)\equiv R(t)$ by examining its RGE:
\begin{mathletters}
\bea
16\pi^2 {d\ln R(t)\over dt} &=& \lt^2 + 3\lb^2 - 3\ltau^2 +
{4\over3} g_1^2 - {16\over3} g_3^2 \label{eqrgR} \\
&\sim& \lt^2\left(4 - {3\over R^2}\right) - {16\over3} g_3^2
\label{eqrgRappr}
\eea
\end{mathletters}
where in the last line we have made the rough approximations
$\lt\simeq\lb$ and $g_3^2\gg g_1^2$.
For very small $\lG$ the first term is negligible, and $R(t)$ is
driven purely by the gauge coupling evolution from $R^G=1$ to
$R(\mu_Z)\simeq 2.4$, independent of  $\lG$. For very large $\lG$,
the first term dominates at the beginning, but now $R$ decreases
almost instantly until $R\simeq \sqrt{3/4}$ while $\lt$ decreases
quickly and becomes independent of $\lG$; the subsequent evolution
from these effective initial conditions to $R(\mu_Z)\simeq1.6$ is
therefore also independent of $\lG$. Thus $R(t)$ has a fixed-point
behavior as a function of $\lG$ for both small and large $\lG$. The
values of interest to us, $0.5 \le \lG \le 2.0$, lie in the
intermediate- to large-$\lG$ range, which is why $R(\mu_Z)$ becomes
less sensitive to $\lG$ as $\lG\rightarrow2$.
\item To understand $F(R)$, observe that both $R$ and $\lt$ display
similar fixed-point dependence on $\lG$, so when we eliminate $\lG$
to obtain $\lt(\mu_Z)=F(R)$ we arrive at a roughly linear dependence
of $\lt$ on $R$.
\item The behavior of $F_{b,\tau}^{(\epsilon)}(R)$ follows from
similar reasoning. For small $\lG$, changing $R^G$ via
$\epsilon_b-\epsilon_{\tau}$ has a reasonably large effect on
$R(m_t)$, which requires (in order to match the experimental value)
a moderate change in $\lG$ --- see Fig.~2a. For large $\lG$, the
same change in $R^G$ has only a small effect on $R(\mu_Z)$ due to
the fixed-point behavior, but now this small effect gets magnified
back into a moderate change required in $\lG$. Thus a fixed
$\epsilon_b-\epsilon_{\tau}$ always necessitates roughly the same
change in $\lG$. But the resulting change in $\lt(\mu_Z)$ is tiny
for large $\lG$ or equivalently for small $R$, and that is why
$F_{b,\tau}^{(\epsilon)}(R)$ become small for small $R$. Now
$F_t^{(\epsilon)}(R) + F_b^{(\epsilon)}(R) +
F_{\tau}^{(\epsilon)}(R) = 0$ because when $\epsilon_t = \epsilon_b
= \epsilon_{\tau}\equiv\epsilon$ the final value of $m_t$ must not
depend on $\epsilon$ (this just amounts to a redefinition of $\lG$).
Hence $F_b^{(\epsilon)}(R) = -F_{\tau}^{(\epsilon)}(R) -
F_t^{(\epsilon)}(R) \simeq -F_{\tau}^{(\epsilon)}(R)$.
\item Finally, $F_t^{(k)}$ is by definition equal to $F(R)$, while
$F_b^{(k)}(R)$ and $F_{\tau}^{(k)}(R)$ measure the changes in
$\lt(\mu_Z)$ induced by changing $R$ directly at the top mass scale,
so their behavior follows immediately from the dependence of $\lt$
on $R$ described above.
\end{itemize}

\section{Predictions for the Top Mass}

The final step is the calculation of the position of the pole in the
propagator of the top quark within the 2-Higgs standard model. (Note
that the pole mass is only a parameter in the calculation of
experimental observables; we leave the study of the relation between
$m_t^{\rm pole}$ and the actual ``top mass'' extracted from collider
data to future work.) There are two important radiative corrections
to the pole mass: the usual QCD correction from gluon dressing, and
Yukawa interaction corrections to the top quark propagator and to
the Fermi constant.
The result may be written as
\bea
m_t^{\rm pole} &=& \lt^{\rm 2HSM} \left[{1\over\sqrt{2^{3/2}G_F}}
\left(1+{\Sigma_W(0)\over 2m_W^2}\right)\right]
\left(1 + {\delta m_t\over m_t}
+ \case{1}{2} \delta^t_L + \case{1}{2} \delta^t_R\right)
\left(1 + \delta_{\rm QCD}\right)
\nonumber\\
&=& \lt\,177\GeV\,\left[1+f_t(m_A)\right] .
\label{eqtpole}
\eea
In the last line we have substituted $\lt^{\rm 2HSM} \equiv \lt$ and
$1+\delta_{\rm QCD}\equiv 1 + 5\alpha_3/3\pi -
(8\alpha_3/4\pi)\ln (m_t/\mu_Z) \simeq 1.015$ when $\mu_Z = m_Z$,
and
defined $f_t(m_A)\equiv \Sigma_W(0)/2m_W^2 + \delta m_t/m_t +
\case{1}{2} \delta^t_L + \case{1}{2} \delta^t_R$. One should not
confuse the pole mass in the above equation and the Euclidean pole
mass defined in Eq. (15): the one just above represents the real
pole of the
propagator at timelike momenta calculated in perturbative QCD,
while the Euclidean pole mass does not actually correspond to any
pole in the
propagator (for obvious reasons).
 Notice also that the above
top pole mass, which is scheme independent, has been written in
terms
 of $\overline {\rm DR}$ quantities. The dominant
1-loop correction to the muon decay constant (used to define $G_F$)
is taken into account via $247\GeV = 1/\sqrt{2^{1/2}G_F} =
v\,\left[1 - \Sigma_W(0)/ 2m_W^2\right]$ where $\Sigma_W(0)$ is the
top contribution to the self energy of the $W$ at zero momentum. The
wavefunction renormalization of the top quark propagator is given by
$\case{1}{2}\delta^t_L$ and $\case{1}{2}\delta^t_R$, while the mass
renormalization $\delta m_t$ is $\mu_Z$-independent  and vanishes in
the 't Hooft-Feynman gauge we employ. The complete expression for
$f_t$ is given in the appendix; to a good approximation, $f_t \simeq
0.04 + 0.003 \ln m_A/\mu_Z$.

The complete form of our prediction for $m_t$ may be obtained by
substituting the expressions for $R$, $\epsilon_n$, and $k_\nu$ into
Eq.~(\ref{eqtsol}) and inserting the result into
Eq.~(\ref{eqtpole}).
To untangle the various dependences of the prediction on the SUSY-
and GUT-scale parameters, we divide the various contributions into
three classes, and discuss each separately before combining them
into a prediction.

First, let us {\it ignore} both the (potentially large) finite
$\delta m_b$ corrections discussed in Secs.~III and VI as well as
GUT-scale
thresholds, and concentrate
on the logarithmic SUSY-scale threshold corrections.
Varying only the superpartner masses affects the predictions in
three ways: through $k_\nu$, through $\epsilon_i$, and (as remarked
above) through the $\alpha$-independent shift in all the $T_\alpha$.
The latter effect (which actually shifts all GUT-scale masses
together) is shown below to be small. The first two can be
significant, but {\it only} when they serve to {\it increase} $m_t$.
For
example, if $\alpha_3(m_Z) = 0.11$, $m_b(m_b) = 4.2\GeV$ and $m_A =
90\GeV$, then when all superpartners have a mass equal to $m_Z$ we
predict $m_t = 167\GeV$. By
allowing the various $\{m_a\}$ to vary between $m_Z$ and
$3\TeV$, we obtain $m_t$ as high as 195 GeV but only as low as 163
GeV. To understand this fact, recall that $F_3^{(\epsilon)}(R) \gg
F_2^{(\epsilon)}(R) \sim F_1^{(\epsilon)}(R)$ so we may neglect
$\epsilon_{1,2}$ relative to $\epsilon_3$. In the three $k_\nu$ the
dominant terms are those proportional to $y_t$, $y_b$ and
$\alpha_3$.  Therefore, for fixed $\alpha_3(m_Z)$  the top mass
prediction depends predominantly (to within a few GeV) on the
squark, gluino and higgsino masses rather than the slepton, wino and
bino masses. Keeping only these terms, we find the approximate
formula
\bea
\lt &\simeq& F +
{\delta\alpha_3\over\alpha_3}
\left[{1\over2}F_3^{(\epsilon)}{\alpha_G\over\alpha_3}\right] -
(t_{\sq}+t_{\sqrt{2}\gl}) \left[2
F_3^{(\epsilon)}{\alpha_G\over4\pi}
\right] +
t_{\sq,\gl}\left[{8\over3}{\alpha_3\over4\pi}(F+F_b^{(k)})\right]
\nonumber \\
&\phantom{=}&+
t_{\sq,\hi}\left[{1\over2}{y_t\over4\pi}(3F+F_b^{(k)}) +
  {1\over2}{y_b\over4\pi}(F+3F_b^{(k)})\right],
\label{lamtappr}
\eea
where the $R$ dependence is implicit and we have used $F_t^{(k)} =
F$. As it turns out, the positive terms proportional to
$t_{\sq,\gl}$ and $t_{\sq,\hi}$ (from wavefunction renormalization
in the $k_\nu$ SUSY threshold corrections) are always larger in
magnitude than the negative terms proportional to $t_{\sq}$ and
$t_{\sqrt{2}\gl}$ (from $\epsilon_3$ SUSY threshold corrections to
the QCD gauge coupling). Thus---if the $\delta m_b$ corrections are
ignored---heavy superpartners inevitably lead to
a heavier top.

The top mass prediction is further influenced \cite{aszpred} by
$\alpha_3(m_Z)$, which changes not only the value of $g_3$ used in
the running of the Yukawa couplings but also the value of $R$ as a
function of $m_b$. Finally, the Higgs mass parameter $m_A$ enters
into the matching between $R$ in the low-energy no-Higgs standard
model and the $R$ in the 2-Higgs standard model, and into the
expression for the pole mass of the top quark. The resulting
dependence of $m_t^{\rm pole}$ on $m_{\sq}$, $m_{\gl}$, $\mu \sim
m_{\hi}$,
$m_A$, $m_b$ and $\alpha_3(m_Z)$ is somewhat complicated, so we
display it in two complementary ways. First, we list in Table I the
predictions of our complete expressions for various choices of the
parameters, again omitting the $\delta m_b$ corrections of Sec.~III.
(Such an omission is clearly unjustified for many of the parameters
chosen for this table---it is only intended to illustrate the
conclusions of the previous paragraph.) In this table we have used
as average values $m_{\wi} = m_{\bi} =
m_{\sl} = 500 \GeV$.
Second, we can approximate $F$, $F_b^{(k)}$ and $F_3^{(\epsilon)}$
as linear functions of $R$, and further approximate $y_t\simeq
F^2/4\pi$ and $y_b\simeq 0.87 y_t$. For a given set of values for
$\{m_A,\alpha_3(m_Z),m_b\}$ we obtain an expression
\beq
m_t = m_t^0 - c_3 (t_{\sq}+t_{\sqrt{2}\gl}) + c_{\sq,\gl}
t_{\sq,\gl} +
c_{\sq,\hi} t_{\sq,\hi}.
\label{eqmtappr}
\eeq
The values of the $c_i$ are given in Table II, and the resulting top
mass values are accurate to within $\sim\pm 5\GeV$ when the various
masses are varied between $m_Z$ and 3 TeV.

%\mediumtext
\begin{table}
\caption{The top quark pole mass predictions for a range of values
of the parameters. All masses are in GeV, and we have used set
$m_{\wi} = m_{\bi} = m_{\sl} = 500 \GeV$, though the results are
almost independent of these masses. We have not included the $\delta
m_b$ corrections which are large in some cases considered in this
table.}
\begin{tabular}{rrrcccccccc}
\multicolumn{3}{c}{$m_A$}
 &90 GeV&1 TeV&90 GeV&1 TeV&90 GeV&1 TeV&90 GeV&1 TeV\\
\multicolumn{3}{c}{$\alpha_3(m_Z)$}
 &$0.11$&$0.11$&$0.12$&$0.12$&$0.11$&$0.11$&$0.12$&$0.12$\\
\multicolumn{3}{c}{$m_b(m_b)$}
 &$4.4\GeV$&$4.4\GeV$&$4.4\GeV$&$4.4\GeV$&$4.0\GeV$&$4.0\GeV$
&$4.2\GeV$&$4.2\GeV$\\
\tableline
$m_{\sq}$&$m_{\gl}$&$m_{\hi}$&&&&&&&\\
$90$&$90$&$90$&$157$&$164$&$174$&$179$&$172$&$177$&$180$&$184$\\
$90$&$90$&$3000$&$166$&$172$&$183$&$188$&$181$&185
&$188$&(\tablenotemark[1])\\
$90$&$3000$&$90$&$161$&$168$&$181$&$186$&$177$&$182$
&$186$&(\tablenotemark[1])\\
$90$&$3000$&$3000$&$169$&$177$&$189$&$195$&$185$
&$190$&(\tablenotemark[1])&(\tablenotemark[1])\\
$3000$&$90$&$90$&$173$&$181$&$194$&(\tablenotemark[1])&
$189$&(\tablenotemark[1])&(\tablenotemark[1])&(\tablenotemark[1])\\
$3000$&$90$&$3000$&$177$&$184$&
$197$&(\tablenotemark[1])&$193$&(\tablenotemark[1])&
(\tablenotemark[1])&(\tablenotemark[1])\\
$3000$&$3000$&$90$&$158$&$167$&$181$&$187$&$178$&$183$&$188$&$193$\\
$3000$&$3000$&$3000$&$162$&$171$&$184$&$190$&$181$&$186$
&$191$&(\tablenotemark[1])\\
\end{tabular}
\tablenotetext[1]{For these parameter values, $\lG>2$.}
\end{table}

%\mediumtext
\begin{table}
\caption{Coefficients of the approximate formula for the top mass,
Eq.~(\protect\ref{eqmtappr}), which is accurate to $\sim\pm 5\GeV$
when the various masses are varied between $m_Z$ and 3 TeV.
}
\begin{tabular}{ccccccccc}
$m_A$ &90 GeV&1 TeV&90 GeV&1 TeV&90 GeV&1 TeV&90 GeV&1 TeV\\
$\alpha_3(m_Z)$
&$0.11$&$0.11$&$0.12$&$0.12$&$0.11$&$0.11$&$0.12$&$0.12$\\
$m_b(m_b)$
&$4.4\GeV$&$4.4\GeV$&$4.4\GeV$&$4.4\GeV$&$4.0\GeV$&$4.0\GeV$
&$4.2\tablenotemark[1]\GeV$&$4.2\tablenotemark[1]\GeV$\\
\tableline
$m_t^0\,(\GeV)$ &$154$&$162$&$173$&$179$&
$170$&$177$&$179$&$185$\\
$c_3\,(\GeV)$ &$4.8$&$4.3$&$4.2$&$3.8$&
$3.8$&$3.4$&$3.8$&$3.3$\\
$c_{\sq,\gl}\,(\GeV)$ &$8.6$&$8.0$&$7.9$&$7.4$&
$7.4$&$6.9$&$7.4$&$6.9$\\
$c_{\sq,\hi}\,(\GeV)$ &$3.9$&$3.8$&$3.8$&$3.8$&
$3.8$&$3.7$&$3.8$&$3.7$
\end{tabular}
\tablenotetext[1]{For $m_b(m_b) = 4.0\GeV$ and $\alpha_3(m_Z)=0.12$,
most parameter choices lead to $\lG>2$.}
\end{table}

Next, we consider the finite corrections to the $b$ quark mass as
discussed in Sec.~III. As we saw, these are small only when the
squared squark masses are much greater than $m_{\gl} \mu$ and $\mu
A_t$.
In this case, choosing for concreteness $m_{\hi} \sim \mu = 100
\GeV$, $m_{\gl} = 300\GeV$,
$m_{\wi} = 100\GeV$, $m_{\sq} = m_{\sl} = 1000\GeV$ and $m_A =
1000\GeV$, and considering several values of $\alpha_3(m_Z)$, we
obtain the predictions shown in Fig.~4 as solid lines. The pole mass
of the top quark is shown as a function of the running $m_b$
parameter (discussed in Sec.~V) indicated on the upper horizontal
axis. The allowed range for $m_b$ according to Eq.~(\ref{eqRpred}),
using $\alpha_3(m_Z) = 0.12$, is also shown on this axis. We learn
that if the SUSY parameters are sufficiently hierarchical that
$\delta m_b$ corrections may be neglected, then the top quark is
predicted to be heavier than $\sim 175\GeV$ for $\alpha_3(m_Z) >
0.115$. It should be remembered that
the prediction for $\alpha_3(m_Z)$ without GUT-scale threshold
corrections is around 0.124, so values much lower than this
correspond to large GUT-scale corrections to $g_3$. Furthermore, we
find that perturbative Yukawa unification demands
$\alpha_3\lesssim0.125$. These last two observations are in rough
agreement with previous authors \cite{cpw,landp}. If it turns out
that $\delta m_b$ is significant, our predictions may change
considerably and become highly dependent on the SUSY parameters. If
$\delta m_b>0$, then $k_b'>0$, and since $F_b^{(k)}(R) > 0$ the top
mass prediction can only {\it increase}; that is, either the change
is small and the top stays near its maximal value of $\sim
180-190\GeV$, or else the change is too big and the corresponding
SUSY parameters are excluded (see Sec.~VI). On the other hand, if
$\delta m_b<0$ the top mass prediction can be significantly reduced.
We show in Fig.~4 the predictions that result from a light squark
spectrum, namely $m_{\hi} \sim \mu = 250 \GeV$, $m_{\gl} = 300\GeV$,
$m_{\wi} = 100\GeV$, $m_{\sq} = m_{\sl} = 400\GeV$ and $m_A =
400\GeV$. The appropriate horizontal axis is now the lower one,
which is obtained from the upper axis by $m_b \rightarrow m_b
+\delta m_b$. (Since $\delta m_b$ depends on $\tan\beta$ and
thereby on $m_t$, we use the central prediction of $m_t$ in the
figure as a rough guide in rescaling the horizontal axis. Also, we
have checked that the fit in Eq.~(\ref{eqfitzero}) is still
reasonably valid
for these low values of $m_t$.) As is evident from the rescaled
bounds on $m_b$ shown in the figure, the top mass prediction now
would encompass essentially all the experimentally allowed range. In
other words, we obtain a meaningful prediction only if the squarks
are much heavier than  $m_{\gl} \mu$ and $\mu A_t$ or if we {\it
know} that $\delta m_b>0$.  We will discuss whether these
requiremnts are likely to be satisfied elsewhere \cite{newradsymm}.

\begin{figure}
\leavevmode
\epsfxsize=5.0in \epsfbox[-20 200  500 500]{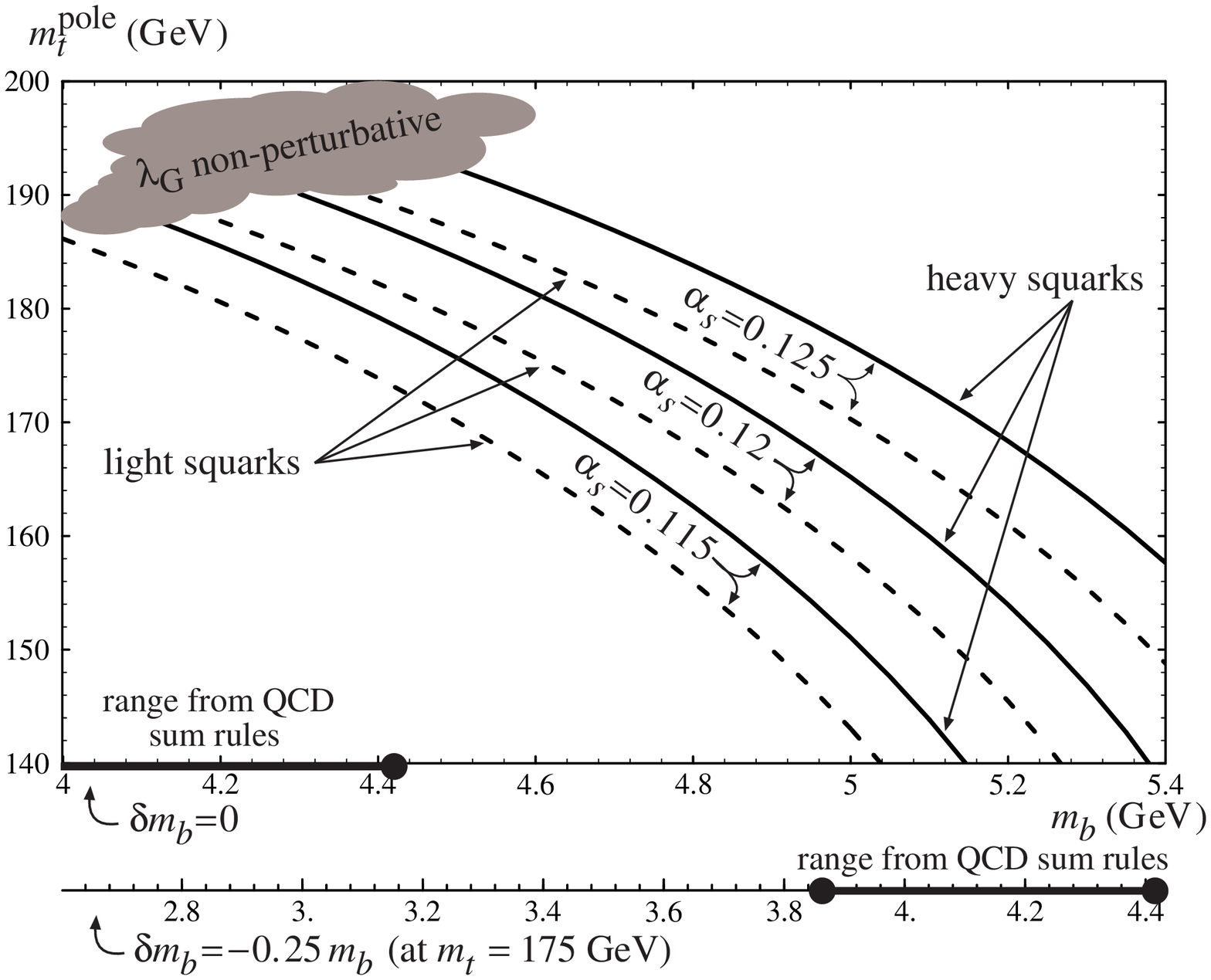}
\begin{quote}
{\small
FIG.~4. Our predictions for
$m^{\rm pole}_t$ without superheavy corrections using two
qualitatively-different
superpartner spectra, specifically $m_{\hi} \sim \mu = 100 \GeV$,
$m_{\gl} = 300\GeV$,
$m_{\wi} = 100\GeV$, $m_{\sq} = m_{\sl} = 1000\GeV$ and $m_A =
1000\GeV$ for the ``heavy squarks" case, and $m_{\hi} \sim \mu = 250
\GeV$, $m_{\gl} = 300\GeV$,
$m_{\wi} = 100\GeV$, $m_{\sq} = m_{\sl} = 400\GeV$ and $m_A =
400\GeV$ for ``light squarks.'' The ``cloud'' indicates the region
where
$\lG>2$. These predictions carry estimated uncertainties of $\sim\pm
5\GeV$ from various approximations and from the GUT-scale thresholds
discussed in the text.
Also shown are the estimated allowed mass ranges for the running
parameter $m_b$ as extracted in Sec.~V.}
\end{quote}
\end{figure}

It is important to note that our prediction for $m_t$, for given
experimental inputs and for $\delta m_b\simeq 0$,  is
considerably larger (by 10 to 20 GeV) than has been previously
obtained using
lowest order analyses \cite{als,alsalt}. Carena, Pokorski and Wagner
\cite{cpw} have briefly considered the condition $\lt=\lb=\ltau$ as
a particular case of GUT boundary conditions, and employed 2-loop
RGEs to numerically obtain top mass values with which we agree in
the minimal scenario. They do not, however, describe the complete
dependences on $m_b$ and $\alpha_3$ nor do they address the question
of $\delta m_b$ or of superheavy corrections. Finally, we reach
different conclusions
about the dependence on the superpartner spectrum.

We turn now to the threshold corrections which may be present at the
GUT scale. These fall into three classes, corresponding to the three
terms that make up $\epsilon_{t,b,\tau}$ in Eqs.~(\ref{eqeps}): the
splitting of the $\underline{16}_3$ and of the $\underline{10}_H$
which contribute in proportion to $y_G$, the splitting of the the
superheavy members of the gauge $\underline{45}$ which contributes
in proportion to $\alpha_G$, and any other superheavy multiplets
which may couple to the $\underline{10}_H$ or the
$\underline{16}_3$. In addition, there is the shift in the
prediction of the overall superheavy mass scale which occurs if the
superpartner mass splittings change the scale at which $g_1$ and
$g_2$ meet to $M_G^{\prime}$. This shift could also result from
GUT-scale threshold corrections to $g_1$ and $g_2$. In any case,
such a shift induces an effective GUT threshold correction given by
$\Delta\epsilon_{\nu} = {1\over 4\pi} \ln (M_G^{\prime}/M_G)
\sum_\alpha
(K_{\nu}^\alpha y_G + L_{\nu}^\alpha \alpha_G + \sum_\sigma K_{\nu
\sigma}^{\prime\alpha} y_\sigma)$ for $\nu=t,b,\tau$. After summing
over all superheavy particles $\alpha$, the last term in
$\Delta\epsilon_{\nu}$ becomes equal for all $\nu$ since it only
involves complete SO(10) multiplets; therefore we may drop it. The
first two terms are ${1\over 4\pi} \ln
(M_G^{\prime}/M_G)\left[(7,6,7)y_G -
({223\over10},{227\over10},{267\over10})\alpha_G\right]$, which
become \cite{pscomment} ${1\over 4\pi} \ln
(M_G^{\prime}/M_G)\left[(0,-1,0)y_G -
(0,{2\over5},{22\over5})\alpha_G\right]$ after subtracting out an
irrelevant constant. Then an $M'_G$ anywhere between
$1.3\times10^{15}\GeV$ and $1.3\times10^{17}\GeV$ changes the top
mass by less than 2\%.

The threshold corrections proportional to $y_G$ arise from the
splitting of the $\underline{16}_3$ into the right-handed neutrino
and the rest of the standard-model matter fields, and of the
$\underline{10}_H$ into the superheavy Higgs triplets and the
standard-model Higgs doublets.
Since the right-handed neutrino is but a small part of the
$\underline{16}_3$, its contribution is small, as we saw above in
$\Delta\epsilon_{\nu}$, and since its mass is expected to be at or
below the GUT scale it can only raise the mass of the top. The
splitting of the $\underline{10}_H$ is dominated by the large
hierarchy between the doublets and the triplets; in the language of
the Pati-Salam subgroup $G_{\rm PS}\equiv \rm SU(2)_L\times
SU(2)_R\times SU(4)$ of SO(10), we write
$\underline{10}_H\rightarrow (2,2,1)+(1,1,6)$. But in fact any
complete multiplet of $G_{\rm PS}$ will contribute equally to
$\lambda_t$, $\lambda_b$ and $\lambda_{\tau}$, so the large
splitting between the  $(2,2,1)$ and the $(1,1,6)$ does not generate
any threshold corrections. Furthermore, the $(1,1,6)$ decomposes
into a $\underline{3}+\underline{\overline{3}}$ of color SU(3),
which as it turns out {\it each} contribute equally to the three
Yukawa couplings (assuming they are heavier than the right-handed
neutrino), so no threshold contributions result from their possible
splitting.

Threshold corrections proportional to $\alpha_G$ would be generated
by splittings amongst the superheavy fields in the gauge
$\underline{45}$ as well as amongst the members of the
$\underline{10}_H$ and $\underline{16}_3$. Only splittings of
$\lb^G$ from $\ltau^G$ are not largely suppressed by the fixed-point
behavior, as discussed above. The dominant contributions to such
splittings come from three multiplets
$\phi_{1,2,3}\in\underline{45}$, having masses $M_{1,2,3}$
respectively, which transform as $\phi_1 \sim (2,3,{1\over3})$,
$\phi_2 \sim (2,3,-{5\over3})$ and $\phi_3 \sim (1,3,{4\over3})$
under
${\rm SU(2)\times SU(3)\times U(1)_Y}$. They yield $\epsilon_b -
\epsilon_\tau = (\alpha_G/\pi)\left[\ln(M_1/M_2) -
\ln(M_3/M_G)\right]$. Even in the somewhat extreme case of $M_2 \sim
10 M_1 \sim 10 M_3 \sim 100 M_G$, the change in $m_t$ is only 2 to 6
GeV.

Finally, there will in general be threshold effects from various
other couplings to the
$\underline{16}_3\,\underline{10}_H\,\underline{16}_3$ vertex. These
obviously depend upon the Higgs content of the theory, and cannot be
estimated without a concrete model. In particular, if some large
multiplet such as a $\underline{126}_H$ has a large coupling to the
$\underline{16}_3$ and is far from degenerate in mass, then large
threshold corrections could result; in that case, the Yukawa
couplings would not in effect be unified. Since we cannot rule out
such a possibility (even after we impose the restriction that no
Landau poles be encountered within, say, and order of magnitude of
$M_G$), our predictions will only be valid for models in which
either the couplings of all multiplets (except the
$\underline{10}_H$) to the $\underline{16}_3$ are small, or such
multiplets are practically unsplit, or they make equal contributions
to $\lb$ and $\ltau$. For this last claim we have used the fact that
$|F_b^{(\epsilon)}(R) + F_{\tau}^{(\epsilon)}(R)| =
|F_t^{(\epsilon)}(R)| \ll 1$: any threshold effects that are equal
for $\lb$ and $\ltau$ or that affect only $\lt$, are greatly
suppressed by the fixed-point behavior. This applies, for example,
to any multiplet that only corrects the  $\underline{10}_H$ leg of
the vertex, since it could only distinguish the doublet that couples
to down-type fields from the doublet that couples to up-type fields
but could not distinguish the $b$ from the $\tau$. Even a 30\%
threshold correction of this sort would affect the top mass
prediction by less than 3\%.

In summary, while we cannot eliminate the possibility of large
GUT-scale threshold corrections to our predictions, we have shown
that all those corrections which are generic to SO(10) SUSY GUTs are
not expected to change $m_t$ by more than a few GeV.

\section{Extensions}

The analysis of Secs.~II through VIII has actually assumed that the
two light Higgs doublets lie entirely within a single irreducible
representation of SO(10). To what extent does the analysis, and the
resulting top quark mass predictions, remain valid when the Higgs
doublets have components in other irreducible representations? In
general there will be a set of mixing angles $\{\theta_{U,i}\}$ and
$\{\theta_{D,i}\}$ describing the components of the two light Higgs
doublets, $H_U$ and $H_D$, in various SO(10) multiplets
($\underline{10}_H$, $\underline{126}_H$, or some other
representation) labeled by $i$. Suppose for now that the third
generation masses are generated by a set of Yukawa interactions
which are all of the form
$\underline{16}_3\,\underline{10}_{H,i'}\,\underline{16}_3$, where
the $\{i'\}$ are a subset of the $\{i\}$. In this case the GUT
boundary condition $\lb=\ltau$ will occur regardless of the values
of $\{\theta_{U,i}\}$ and $\{\theta_{D,i}\}$; on the other hand,
$\lt \neq \lb$ if $\theta_{U,i'}\neq\theta_{D,i'}$ for any $i'$. The
analysis of this paper still applies to this situation provided an
additional term $\Delta\epsilon_t$ is added to $\epsilon_t$ to
reflect this change in boundary condition. As we saw above, the
fixed-point behavior implies a considerable insensitivity to
$\epsilon_t$ (see the corresponding sensitivity function
$F^\epsilon_t$). Typically $\Delta\epsilon_t\sim
\theta_{U,i}-\theta_{D,i}$ so for
$\theta_{U,i}-\theta_{D,i}\lesssim0.3$ the shift in the top mass is
less than 5 GeV. (This is not necessarily an uncertainty: in a given
model the mixings are computable and so is the top mass shift.) This
shows that our results apply even when several
$\underline{16}_3\,\underline{10}_H\,\underline{16}_3$ interactions
contribute to the third generation masses and $H_U$ and $H_D$ have
sizeable components in different $\underline{10}_H$ multiplets.
A particularly interesting subclass of the above models contains
just one pair of $\underline {10}$'s, of which only one,
$\underline{10}_1$,
couples to the $\underline {16}_3$. The $\underline{10}_2$
is introduced to make the triplets in $\underline {10}_1$
heavy. This is achieved while keeping the doublets light via the
coupling $\underline{10}_1 \underline{45}_H
\underline{10}_2$, provided the $\underline{45}_H$
gets a VEV in the $B-L$ direction \cite{dimwil}. A more detailed
discussion of a model of this type will be given elsewhere
\cite{newradsymm}.

The restrictions on the values of $\{\theta_{D,i'}\}$ are stricter
when
$i'$ refers to a $\underline{126}_{H,i'}$ which couples via
$\xi_{i'}\underline{16}_3
\,\underline{126}_{H,i'}\,\underline{16}_3$, because this leads to
$\lb\neq\ltau$ at the GUT scale. Using our analysis
then requires an additional contribution to
$\epsilon_b-\epsilon_\tau$
of order $(\xi_{i'}/\lambda_G) \theta_{D,i'}$. Shifts in $m_t$ of up
to 10 GeV occur when $(\xi_{i'}/\lambda_G) \theta_{D,i'}\sim 0.1$.
We conclude that while $H_U$ and $H_D$
must lie predominantly in a single $\underline{10}_H$, they may have
a certain amount of mixing with doublets in other multiplets, where
the allowed mixing may be quite large if the other multiplet is a
$\underline{10}_H$ but must be small if it is a $\underline{126}_H$
with a significant coupling to the third generation.

Another contribution to $\epsilon_{t,b,\tau}$ may arise from
operators of the form $\underline{16}_2\,{\cal O}\,\underline{16}_3$
which mix the second and third generations. Such operators must be
present in order to generate $V_{cb}$; in fact, ${\cal O}$ must have
sufficient structure to give different values for the (2,3) entries
of the up and down Yukawa matrices. This implies a non-universal
contribution to $\epsilon_{t,b,\tau}$. Normally the contribution is
of order $V_{cb}^2\sim10^{-3}$, which hardly affects the top quark
mass prediction. However, if in some scheme (see, for example, the
last reference in \cite{latetextures}) the operator
$\underline{16}_2\,{\cal O}\,\underline{16}_3$ is responsible for
generating a sizable fraction of $m_\mu/m_\tau$ and $m_s/m_b$
[rather than having these generated by the (2,2) entries of the
Yukawa matrices] then the contribution to $\epsilon_{b,\tau}$ could
be large. For example, if there is {\it no} operator of the form
$\underline{16}_2\,{\cal O}\,\underline{16}_2$ to give a mass to the
muon or the strange quark, then we would expect
$\epsilon_b-\epsilon_\tau \sim m_\mu/m_\tau \sim 0.04$, which would
lead to a shift in $m_t$ of $\sim 8\GeV$. This extreme case is
however unnatural since such schemes will not lead to an
understanding of why $V_{cb}^2\sim m_c/m_t \ll m_\mu/m_\tau \sim
m_s/m_b$; in any case, the shift in such schemes is calculable and
does not introduce much uncertainty into the prediction of the top
mass.

\section{Conclusions}

In this paper we have computed the top quark mass in supersymmetric
SO(10) grand
unified theories including all contributions larger than about 5
GeV. The
assumption underlying this computation is that $m_t$, $m_b$ and
$m_\tau$ all
originate from a single Yukawa interaction with an SO(10) multiplet
which  dominantly contains the two low energy Higgs
doublets. While this is a somewhat restrictive assumption, it has
certain virtues. It
seems to us to be the simplest assumption, within grand unified
theories, which
can lead to a top mass prediction in terms of low energy quantities.
In particular it does not involve
any
ansatz about the origin of masses for the lighter two generations.
Furthermore this assumption leads to an almost unique
group-theoretic
structure which
underlies the top mass prediction: except for a narrow range of
(measurable) MSSM parameters, the Yukawa interaction must be of
the form $\underline{16}_3\,\underline{10}_H\,\underline{16}_3$.
The top mass prediction which results from
the unification of the three third-generation Yukawa interactions
recalls the prediction of the weak
mixing angle which follows from the unification of the three gauge
couplings. There is however an important difference between these
two
cases, having to do with threshold effects at the SUSY scale.  In
the gauge case these effects consist of a renormalization of {\it
only
one} operator (the gauge kinetic term), and are therefore small as
long as the theory is perturbative. In contrast, there are two
possible Yukawa interaction operators
below the SUSY scale, corresponding to the two Higgs doublets, that
contribute to the mass of each fermion
once these doublets acquire VEVs. Thus a hierarchy of
VEVs could result in large threshold corrections to
the lighter masses without necessarily invalidating perturbation
theory. Such a hierarchy is actually mandated in the SO(10)
unification under our assumptions, and results in another contrast
with previous unification scenarios, specifically the partial Yukawa
unification in SU(5).
A crucial outcome of the GUT boundary condition $\lb=\ltau$ is that
the resulting value of $m_b/m_\tau$ can only be reconciled with
experiment if the top quark Yukawa coupling is large. This is why in
SU(5) and in SO(10) models the top quark is predicted to be heavy.
In the former, the ratio of $\lt/\lb$ at the GUT scale is an
arbitrary parameter, which thus precludes a prediction of $m_t$.
The SO(10) GUT boundary condition fixes $\lt/\lb$ to be unity;
however,
we are forced to take
$v_U/v_D\gg 1$ to account for $m_t/m_{b,\tau}\gg 1$, and hence large
SUSY threshold effects introduce a new parameter $\delta m_b$ into
the top mass prediction. This feature of SO(10) is an
improvement over SU(5): now the undetermined parameter
is a ratio of observable mass parameters, which
is measurable in principle by future experiments, rather than a
ratio of
VEVs which will be more difficult to deduce.
Furthermore, if we assume a hierarchical structure in the SUSY
spectrum (a discussion of such a structure will be presented
elsewhere \cite{newradsymm}) in which the sfermions are considerably
heavier than gauginos
and higgsinos, there are indeed no large threshold effects and the
top
mass is sharply predicted.

%understanding of SO(10) Yukawa unification.

With these caveats in mind, and choosing the hierarchical spectrum,
our basic results are shown in Fig.~3 and Table
I, and are approximated by Eq.~(\ref{eqmtappr}) and Table II. These
show the dependence of the predicted top quark pole mass on
$\alpha_3$ and $m_b$ and
the dominant superpartner spectrum effects. Unless future
experiments
find a value for $\alpha_3$ below the range shown in the figure, the
hierarchical spectrum predicts a top quark heavier than $\sim
170\GeV$ to within 5
GeV or so. (Of course, various uncertainties could pile up and
result in a lower top mass, but this is unlikely.) A smaller
experimental value
for $m_t$ would indicate that $\delta m_b$ is considerable;
measuring the various MSSM mass parameters could then test this
SO(10) unification scenario as outlined at the end of Sec.~VI. More
generally, our prediction reads $m_t^{\rm pole} =
\lt\,(177\GeV)\,[1+f_t(m_A)]$ where $\lt = F(R) + \sum_n
F_n^{(\epsilon)}(R) \epsilon_n + \sum_\nu F_\nu^{(k)}(R) k_\nu$,
$F(R)$ and the sensitivity functions $F_n^{(\epsilon)}(R)$ and
$F_\nu^{(k)}(R)$ are shown in Fig.~2, the $\epsilon_n$ and the
$k_\nu$ are given in Eqs.~(\ref{eqeps1}-\ref{eqktau}), and $R  =
\left[m_b^{\msbar}(4.1\GeV)/m_\tau\right] \left(\eta_\tau /
\eta_b\right) \left[1-\alpha_3(\mu_Z)/3\pi\right] \left[1 +
f_R(m_A)\right]$. In any specific MSSM-based GUT model in which
$\lt^G\simeq\lb^G\simeq\ltau^G$, both the possible contributions of
the diagrams in Fig.~1 and any further deviations from the equality
of GUT-scale Yukawa couplings can be calculated and inserted into
the above expressions. These then yield an analytic prediction of
the top mass to within a few GeV---which is more than sufficient for
comparison with top mass measurements likely to be made in the near
future.

{\bf Note added:} After this manuscript was completed, we received a
paper by M.~Carena, M.~Olechowski, S.~Pokorski and C.E.M.~Wagner
\cite{newcopw}, in which some of the same issues as in our work are
addressed and studied in a particular context, namely that of
universal soft masses at the GUT scale. Many of these issues are
discussed in detail in our Ref.~\cite{newradsymm}. A few have also
appeared in Ref.~\cite{newas}.

\acknowledgments

We would like thank B. Ananthanarayan for useful discussions and G.
Anderson for comparisons with his numerical results, which will
appear in the last reference in \cite{latetextures}.  This work
was supported in part by the Director, Office of Energy
Research, Office of High Energy and Nuclear Physics, Division of
High
Energy Physics of the U.S. Department of Energy under Contract
DE-AC03-76SF00098, and in part by NSF grant PHY-90-21139.

\appendix
\section*{Useful Functions}
We collect below various useful functions needed for calculating
corrections to $m_b$, $R$ and $m_t^{\rm pole}$. The integral which
arises in calculating the finite 1-loop corrections to the $b$ mass
in Fig.~1 is given by
\bea
I(x,y,z) &\equiv& \int_0^\infty {u\,du\over (u+x)(u+y)(u+z)}
\nonumber\\
&=& - {x\,y\ln x/y + y\,z\ln y/z + z\,x\ln z/x \over
[(x-y)(y-z)(z-x)]}.
\label{ifn}
\eea
In some of the SUSY threshold corrections we need the function
\bea
G_1(x,y) &\equiv& \int_0^1 u \ln\left[(1-u)\,x + u\,y\right]\,du
\nonumber\\
&=& {1\over(x-y)^2}
\left\{x^2 \left(\ln{x} - \case{1}{2}\right) -
y^2 \left(\ln{y} - \case{1}{2}\right) \right.
\nonumber\\
&\phantom{=}& \phantom{{1\over(x-y)^2}}\left.
\vphantom{y^2} - 2 y \left[x\left(\ln x - 1\right) -
y \left(\ln y - 1\right)
\right]\right\}.
\label{gone}
\eea
When matching $R$ in the 2HSM and the 0HSM, we use both $I(x,y,z)$
and $G_1(x,y)$ through the function
\bea
16\pi^2 f_R(m_A) &=& (\lb^2-\ltau^2)\,
G_1\left({m_A^2\over\mu_Z^2},0\right) +
\case{1}{2}\lt^2\,G_1\left({m_W^2\over\mu_Z^2},
{m_t^2\over\mu_Z^2}\right)
\nonumber\\
&\phantom{=}&\vphantom{x}
+ \case{1}{2}\lb^2\,G_1\left({m_A^2+m_W^2\over\mu_Z^2},
{m_t^2\over\mu_Z^2}\right) -
\case{1}{2}\ltau^2\,
G_1\left({m_A^2+m_W^2\over\mu_Z^2},0\right)
\nonumber\\
&\phantom{=}&\vphantom{x}
+ \lt^2\,m_A^2\,I(m_A^2+m_W^2,m_t^2,m_W^2).
\label{freqn}
\eea
Finally, for the top quark pole mass, the relevant function is
\beq
f_t(m_A) = {\Sigma_W(0)\over 2m_W^2} + {\delta m_t\over m_t}
+ \case{1}{2} \delta^t_L + \case{1}{2} \delta^t_R
\label{fteqn}
\eeq
where
\bea
16\pi^2 {\Sigma_W(0)\over 2m_W^2} &=&
3\lt^2 \left(\ln {m_t\over\mu_Z} - {1\over4}\right),
\label{fndelv}\\
\phantom{16\pi^2} {\delta m_t\over m_t} &=& 0,
\label{fndelmt}\\
16\pi^2 \delta^t_L &=&
\lt^2\,G_2\left({m_t^2\over\mu_Z^2},{m_Z^2\over\mu_Z^2}\right) +
\lb^2\,G_3\left({m_t^2\over\mu_Z^2},{m_A^2\over\mu_Z^2}\right),
\label{fndelL}\\
16\pi^2 \delta^t_R &=&
\lt^2\,G_2\left({m_t^2\over\mu_Z^2},{m_Z^2\over\mu_Z^2}\right) +
\lt^2\,G_3\left({m_t^2\over\mu_Z^2},{m_Z^2\over\mu_Z^2}\right),
\label{fndelR}
\eea
and
\bea
G_2(x,y) &\equiv& \int_0^1 u \ln\left[(1-u)^2\,x + u\,y\right]\,du,
\label{fngtwo}\\
G_3(x,y) &\equiv& \int_0^1 u \ln\left|u (1-u)\,x - u\,y\right|\,du.
\label{fngthree}
\eea


\begin{references}
\bibitem{lhaddr} E-mail: hall\_lj@theorm.lbl.gov
\bibitem{rraddr} Present address: Physics Department, Rutgers
University, Piscataway, NJ  08855;
e-mail: rattazzi@physics.rutgers.edu
\bibitem{usaddr} Present address: Physics Department, Stanford
University, Stanford, CA 94305; e-mail: sarid@squirrel.stanford.edu
\bibitem{ssq} H. Georgi, H. Quinn, and S. Weinberg, Phys.\ Rev.\
Lett.\ {\bf 33}, 451 (1974); S. Dimopoulos, S. Raby, and F. Wilczek,
Phys.\ Rev.\ D {\bf 24}, 1681 (1981); S. Dimopoulos and H. Georgi,
Nucl.\ Phys.\ {\bf B193}, 150 (1981); L. Iba\~{n}ez and G.G. Ross,
Phys.\ Lett.\ B {\bf105}, 439 (1981); M.B. Einhorn and D.R.T. Jones,
Nucl.\ Phys.\ {\bf B196}, 475 (1982); W.J. Marciano and G.
Senjanovic, Phys.\ Rev.\ D {\bf 25}, 3092 (1982).
\bibitem{caveat} At higher order there are dependences on the
spectrum of the
theory; however it is remarkable how insensitive $\sin^2\theta_W$ is
to the details of this spectrum.
\bibitem{gg} H. Georgi and S.L. Glashow, Phys.\ Rev.\ Lett.\ {\bf
32}, 438 (1974).
\bibitem{ceg} M. Chanowitz, J. Ellis, and M.K. Gaillard, Nucl.\
Phys.\ {\bf B135}, 66 (1978).
\bibitem{twohiggs} L.E. Iba\~{n}ez and C. Lopez, Phys.\ Lett.\ B
{\bf126}, 54 (1983) and Nucl.\ Phys.\ {\bf B233}, 511 (1984); H.
Arason {\it et al.} \cite{alsalt}; L.J. Hall and U. Sarid, Phys.\
Lett.\ B {\bf271}, 138 (1991); S. Kelley {\it et al.} \cite{alsalt}.
\bibitem{gj} H. Georgi and C. Jarlskog, Phys.\ Lett.\ B {\bf86}, 297
(1979).
\bibitem{gn} H. Georgi and D.V. Nanopoulos, Nucl.\ Phys.\ {\bf
B159}, 16 (1979).
\bibitem{hrr} J. Harvey, P. Ramond, and D.B. Reiss, Phys.\ Lett.\ B
{\bf92}, 309 (1980) and Nucl.\ Phys.\ {\bf B199}, 223 (1982).
\bibitem{latetextures} S. Dimopoulos, L.J. Hall, and S. Raby, Phys.\
Rev.\ Lett.\ {\bf 68}, 1984 (1992), Phys.\ Rev.\ D {\bf 45}, 4192
(1992), and Phys.\ Rev.\ D {\bf 46}, R4793 (1992); P. Ramond,
University of Florida Report No.~UFIFT-92-4 (1992); H. Arason, D.
Casta\~{n}o, E.J. Piard, and P. Ramond, Phys.\ Rev.\ D {\bf 47}, 232
(1993); G.F. Giudice, Mod. Phys. Lett. {\bf A7}, 2429 (1992); V.
Barger, M.S. Berger, T. Han, and M. Zralek, Phys.\ Rev.\ Lett.\ {\bf
68}, 3394 (1992); G.W. Anderson, S. Raby, S. Dimopoulos, and L.J.
Hall, Phys.\ Rev.\ D {\bf 47}, 3702 (1993);
G.W. Anderson, S. Dimopoulos, L.J. Hall, S. Raby, and G. Starkman,
Lawrence Berkeley Laboratory Report No.~LBL-33531.
\bibitem{soten} H. Georgi, in {\it Particles and Fields--1974},
edited by C.E. Carlson, AIP Conference Proceedings No.\ 23 (American
Institute of Physics, New York, 1975), p. 575; H. Fritzsch and P.
Minkowski, Ann.\ Phys.\ (N.Y.) {\bf 93}, 193 (1975).
\bibitem{neutmass} S. Dimopoulos, L.J. Hall, and S. Raby, Phys.\
Rev.\
D {\bf 47}, 3697 (1993);  K.S. Babu
and R.N. Mohapatra, Phys.\ Rev.\ Lett.\ {\bf 70}, 2845
(1993).
\bibitem{fixedpt} B. Pendleton and G.G. Ross, Phys.\ Lett.\ B
{\bf98}, 291 (1981); C.T. Hill, Phys.\ Rev.\ D {\bf 24}, 691 (1981).
\bibitem{als} B. Ananthanarayan, G. Lazarides, and Q. Shafi, Phys.\
Rev.\ D {\bf 44}, 1613 (1991); Q. Shafi and B. Ananthanarayan, {\it
Proceedings of the 1991 Trieste Summer School}, edited by E. Gava
{\it et al.} (World Scientific, Singapore, 1992), p. 233 [Bartol
Research Institute Report No.~BA-91-76 (1991)].  \bibitem{alsalt} H.
Arason, D.J. Casta\~{n}o, B.E. Keszthelyi, S. Mikaelian, E.J. Piard,
P. Ramond, and B.D. Wright, Phys.\ Rev.\ Lett.\ {\bf 67}, 2933
(1991); S. Kelley, J.L. Lopez, and D.V. Nanopoulos, Phys.\ Lett.\ B
{\bf274}, 387 (1992); V. Barger, M.S. Berger, and P. Ohmann, Phys.\
Rev.\ D {\bf 47}, 1093 (1993).
\bibitem{sufive} A prediction can be made for the ratio
$m_t/\sin\beta$ in SU(5) but not directly on the top quark mass.
\bibitem{nonsusy} Such a prediction can also be made in
non-supersymmetric SO(10) theories. However, whatever mechanism is
used to rectify the $\sin^2\theta_W$ prediction in the theory will
also affect the top prediction, making it much more model-dependent.
\bibitem{newradsymm} R. Rattazzi and U. Sarid, in
preparation; R. Rattazzi and U. Sarid, to appear in {\it
Proceedings of the Second IFT Workshop on Yukawa Couplings and the
Origins of Mass} (1994); one of the necessary fine-tunings necessary
in such scenarios has also been shown by A.E. Nelson and L. Randall,
Phys.\ Lett.\ B
{\bf 316}, 516 (1993). Previously the issue of radiative symmetry
breaking in such scenarios was considered by G.F. Giudice and G.
Ridolfi, Z.\ Phys.\ C {\bf
41}, 447 (1988); M. Olechowski and S. Pokorski, Phys.\ Lett.\ B
{\bf214}, 393 (1988);  W. Majerotto and B. M\"{o}sslacher, Z.\
Phys.\ C {\bf 48}, 273 (1990); P.H. Chankowski, Phys.\ Rev.\ D {\bf
41}, 2877 (1990); M. Drees and M.M. Nojiri, Nucl.\ Phys.\ {\bf
B369}, 54 (1992); B. Ananthanarayan, G. Lazarides, and Q. Shafi,
Phys.\ Lett.\ B {\bf 300}, 245 (1993).
\bibitem{babubarr} K.S. Babu and S.M. Barr,  Phys.\ Rev.\ D {\bf
48}, 5354 (1993);
\bibitem{rge} D.R.T. Jones, Phys.\ Rev.\ D {\bf 25}, 581 (1982);
J.E. Bj\"{o}rkman and D.R.T. Jones, Nucl.\ Phys.\ {\bf B259}, 533
(1985);
M.E. Machacek and M.T. Vaughn, Nucl.\ Phys.\ {\bf B222}, 83 (1983),
Nucl.\ Phys.\ {\bf B236}, 221 (1984) and Nucl.\ Phys.\ {\bf B249},
70 (1985); see also the appendix in Barger, Berger and Ohmann
\cite{alsalt}.
\bibitem{expdata} P. Langacker and N. Polonsky, Phys.\ Rev.\ D {\bf
47}, 4028 (1993).
\bibitem{cpw} M. Carena, S. Pokorski and C.E.M. Wagner, Nucl.\
Phys.\
{\bf B406} 59 (1993).
\bibitem{taumass} J.Z. Bai {\it et al.} (BES Collaboration), Phys.\
Rev.\ Lett.\ {\bf 69}, 3021 (1992).
\bibitem{banks}  T. Banks, Nucl.\ Phys.\ {\bf B303}, 172 (1988).
\bibitem{mubound} See, for example, F. Zwirner,  in {\it Physics and
experiments with linear colliders}, edited by R.~Orava, P.~Eerola
and M.~Nordberg (World
Scientific, Singapore, 1992), Vol.~I, p.~309 [CERN Theoretical
Report No.~CERN-Th.6357/91], and references therein.
\bibitem{largetanbhiggs} See, for example, J. Gunion, H. Haber, G.
Kane, and S. Dawson, {\it The Higgs Hunter's Guide} (Addison-Wesley,
Reading, MA, 1990).
\bibitem{pdg} K. Hikasa {\it et al.} (Particle Data Group), Phys.\
Rev.\ D {\bf 45}, part II (1992).
\bibitem{sumrules} E.C. Poggio, H.R. Quinn, and S. Weinberg, Phys.\
Rev.\ D {\bf 13}, 1958 (1976); M.A. Shifman, A.I. Vainshtein, and
V.I. Zakharov, Nucl.\ Phys.\ {\bf B147}, 385 and 448 (1979); B.
Guberina, R. Meckbach, R.D. Peccei, and R. R\"{u}ckl, Nucl.\ Phys.\
{\bf B184}, 476 (1981).
\bibitem{gp} H. Georgi and H.D. Politzer, Phys.\ Rev.\ D {\bf 14},
1829 (1976).
\bibitem{gl} J. Gasser and H. Leutwyler,  Phys.\ Rep.\ {\bf 87}, 77
(1982).
\bibitem{aszpred} If for some reason GUT- and Planck-scale
corrections to the gauge couplings are known, then $\alpha_3(m_Z)$
becomes a function of the superpartner spectrum, and in particular
of the higgsino mass. The dependence may be found in J. Hisano, H.
Murayama, and T. Yanagida, Phys.\ Rev.\ Lett.\ {\bf 69}, 1014 (1992)
and Nucl.\ Phys.\ {\bf B402}, 46 (1993); Hall and Sarid
\cite{gravsm}; and Langacker and Polonsky \cite{expdata}.
\bibitem{gravsm} See, for example, L.J. Hall and U. Sarid, Phys.\
Rev.\ Lett.\ {\bf 70}, 2673 (1993); Langacker and Polonsky
\cite{expdata}; and references therein.
\bibitem{landp} P. Langacker and N. Polonsky, University of
Pennsylvania Report No.~UPR-0556-T (1993).
\bibitem{pscomment} The approximate equality of the coefficients of
$y_G$ in the RGE of $\lambda_{t,b,\tau}$ becomes an exact one when
the right-handed neutrino is included in the low-energy theory. This
is a consequence of the Pati-Salam subgroup $G_{\rm PS}\equiv \rm
SU(2)_L\times SU(2)_R\times SU(4)$ of SO(10) which is only broken in
the Yukawa sector by the removal of $\nu_R$. In practice, $G_{\rm
PS}$ is manifested when $y_G > \alpha_G$ in which case the three
Yukawa couplings run almost in parallel.
\bibitem{dimwil} S. Dimopoulos and F. Wilczek, Institute for
Theoretical Physics at Santa Barbara preprint UM HE 81-71 (1981),
unpublished.
\bibitem{newcopw} M.~Carena, M.~Olechowski, S.~Pokorski and
C.E.M.~Wagner, CERN Theoretical Report No.~CERN-Th.7163/94 (1994).
\bibitem{newas} B. Ananthanarayan and Q. Shafi, Bartol Research
Institute Report No.~BA-93-25 (1993).

\end{references}
\end{document}